\documentclass[11pt]{article}
\usepackage[latin9]{inputenc}
\usepackage{geometry}
\usepackage{color}
\usepackage{array}
\usepackage{verbatim}
\usepackage{mathtools}
\usepackage{url}
\usepackage{multirow}
\usepackage{amsmath}
\usepackage{amssymb}
\usepackage{graphicx}
\usepackage{setspace}
\usepackage{esint}
\usepackage[authoryear]{natbib}
\makeatletter

\providecommand{\tabularnewline}{\\}
\@ifundefined{date}{}{\date{}}
\bibpunct{(}{)}{;}{autheryear}{,}{;}

\newcommand{\biblist}{\begin{list}{}
{\listparindent 0.0cm \leftmargin 0.50cm \itemindent -0.50 cm
\labelwidth 0 cm \labelsep 0.50 cm
\usecounter{list}}\clubpenalty4000\widowpenalty4000}
\newcommand{\ebiblist}{\end{list}}

\usepackage{latexsym}
\usepackage{multirow}
\usepackage{psfrag}\usepackage{epsf}\usepackage{enumerate}
\usepackage{url}% not crucial - just used below for the URL 
\usepackage{array}
\usepackage{bm}
\usepackage{multirow}
\allowdisplaybreaks

\newtheorem{lemma}{Lemma}\newtheorem{theorem}{Theorem}\newtheorem{assumption}{Assumption}\newtheorem{remark}{Remark}\newtheorem{example}{Example}

\newcommand{\be}{\begin{equation}}
\newcommand{\en}{\end{equation}}
\newcommand{\bea}{\begin{eqnarray}}
\newcommand{\ena}{\end{eqnarray}}
\newcommand{\ba}{\begin{array}}

\newcommand{\ea}{\end{array}}

\newcommand{\T}{\mathrm{\scriptscriptstyle T}}
\newcommand{\mi}{ {\mathrm{mi}}}

\newcommand{\var}{ {\mathrm{var}}} 
\newcommand{\cov}{ {\mathrm{cov}}}

\newcommand{\bone}{ {\mathbf{1}}}

\newcommand{\bP}{ {\mathbb{P}}}
\newcommand{\E}{ {\mathbb{E}}}
\newcommand{\V}{ {\mathbb{V}}}  
 
\newcommand{\F}{ {\mathcal{F}}}

\newcommand{\de}{ {\mathrm{d}}} 
\newcommand{\sen}{ {\mathrm{sen}}} 

\newcommand*{\indep}{%
  \mathbin{%
    \mathpalette{\@indep}{}%
  }%
}
\newcommand*{\nindep}{%
  \mathbin{%                   % The final symbol is a binary math operator
    \mathpalette{\@indep}{\not}% \mathpalette helps for the adaptation
  }%
}
\newcommand*{\@indep}[2]{%
  \sbox0{$#1\perp\m@th$}%        box 0 contains \perp symbol
  \sbox2{$#1=$}%                 box 2 for the height of =
  \sbox4{$#1\vcenter{}$}%        box 4 for the height of the math axis
  \rlap{\copy0}%                 first \perp
  \dimen@=\dimexpr\ht2-\ht4-.2pt\relax
  \kern\dimen@
  {#2}%
  \kern\dimen@
  \copy0 %                       second \perp
}

\makeatother

\begin{document}
\baselineskip .3in 
\title{\textbf{SMIM: a unified framework of Survival sensitivity analysis
using Multiple Imputation and Martingale }}
\author{Shu Yang\thanks{Department of Statistics, North Carolina State University, NC 27695,
U.S.A. Email: syang24@ncsu.edu},$\ $ Yilong Zhang$^{\dagger}$, Guanghan Frank Liu$^{\dagger}$,
and Qian Guan\thanks{Merck \& Co., Inc., Kenilworth, NJ 07033, USA}$\ \ $\thanks{We thank Dr. Gregory Golm and Dr. Gang Jia for useful suggestions
and discussion. We thank Dr. Yale Mitchel and Dr. Joerg Koglin to
review the IMPROVE-IT analysis results in this paper. Yang is partially
supported by the National Science Foundation grant DMS 1811245, National
Cancer Institute grant P01 CA142538, National Institute on Aging grant
1R01AG066883, and National Institute of Environmental Health Science
grant 1R01ES031651.}}
\maketitle
\begin{abstract}
Censored survival data are common in clinical trial studies. We propose
a unified framework for sensitivity analysis to censoring at random
in survival data using multiple imputation and martingale, called
SMIM. The proposed framework adopts the $\delta$-adjusted and control-based
models, indexed by the sensitivity parameter, entailing censoring
at random and a wide collection of censoring not at random assumptions.
Also, it targets for a broad class of treatment effect estimands defined
as functionals of treatment-specific survival functions, taking into
account of missing data due to censoring. Multiple imputation facilitates
the use of simple full-sample estimation; however, the standard Rubin's
combining rule \textcolor{black}{may over estimate the variance for
inference} in the sensitivity analysis framework. We decompose the
multiple imputation estimator into a martingale series based on the
sequential construction of the estimator and propose the wild bootstrap
inference by resampling the martingale series. The new bootstrap inference
has a theoretical guarantee for consistency and is computationally
efficient compared to the non-parametric bootstrap counterpart. We
evaluate the finite-sample performance of the proposed SMIM through
simulation and an application on a HIV  clinical trial. 
\end{abstract}

\newpage{}

\section{Introduction}

Censored survival outcomes are common in clinical trial research of
chronic diseases, such as respiratory, cardiovascular, cancer, and
infectious diseases. As in the missing data literature, three assumptions
about the censoring mechanism have been proposed: censoring completely
at random (CCAR), censoring at random (CAR), and censoring not at
random (CNAR) \citep{tsiatis2007semiparametric}. Censoring due to
administrative constraints, e.g., the planned end of the study, is
unrelated to the study treatment or the underlying health condition.
Therefore, the event times are likely to be CCAR. On the other hand,
the censored event times due to non-administrative reasons such as
premature dropout are unlikely to be CCAR. For example, subjects may
withdraw from the study because of adverse events. Common survival
analysis methods assume CAR that \textcolor{black}{patients censored
at $t$ and patients uncensored at $t$ with the same past history
have the same distribution of the entire current and future variables.}\textit{}%
\begin{comment}
the censoring process is independent of the event time process conditional
on observed covariates; i.e., the censoring process is explainable
by observed information. {[}\textcolor{black}{wrong?}{]}
\begin{itemize}
\item \textit{The definition of coarsening at random is not correct. See
\citet{gill1997coarsening}. The MAR assumption says that subjects
uncensored at time t and subjects censored at time t with the same
recorded past have the same distribution of the entire current and
future variables (Y, V).}
\end{itemize}
\end{comment}
{} This assumption will be violated if sicker subjects are more likely
to withdraw from the study, even after accounting for their observed
history, leading to CNAR. Unfortunately, the censoring assumptions
are not commonly testable from the observed data \citep{rubin1976inference}.
Inappropriate assumptions may lead to biased and misleading conclusions.
In such settings, regulatory agencies, such as the FDA, and national
research council (NRC, \citealp{NRC2010prevention}) request or recommend
sensitivity analyses to assess the robustness of study conclusions
to unverifiable assumptions. 

In this article, we distinguish different reasons for censoring including
administrative reasons and non-administrative reasons. For the latter,
we consider patient premature dropout, which could be a case of CNAR.
\textcolor{black}{Many sensitivity analysis approaches have been developed
for CNAR survival data. One approach is to specify a range of the
residual dependence of the hazard of censoring times on the event
times for the sensitivity parameter; see, e. g., \citet{andrea2001methods},
\citet{scharfstein2002estimation} and \citet{rotnitzky2007analysis}.}%
\textcolor{black}{{} A different approach is to directly specify pattern
mixture models \citep{little1993pattern} for event times for censored
and uncensored patients and impute the missing outcomes for the censored
subjects. \citet{zhao2014multiple} considered Kaplan-Meier curves
to impute data, which, however, cannot include covariates.} Alternatively,
the $\delta$-adjusted \citep{jackson2014relaxing,lipkovich2016sensitivity}
and control-based \citep{lu2015comparison,atkinson2019reference}
models are flexible to accommodate auxiliary information for sensitivity
analysis of unverifiable missing data assumptions. Due to the transparency,
these models have been widely used in applied statistics to handle
missing data (e.g., \citealp{little2012prevention} and \citealp{ratitch2013missing}).
For generality, we consider a class of $\delta$-adjusted/control-based
Cox models for censoring due to premature dropout, indexed by sensitivity
parameter $\delta$. In $\delta$-adjusted models, $\delta$ is a
parameter comparing the outcome distribution of the subjects after
non-administrative censoring with the outcome distribution of the
same subjects had they remained on study. Although we consider the
two reasons for censoring, our framework extends readily to multiple
reasons by adopting different $\delta$'s for different groups. \textcolor{black}{Control-based
models assume that the hazard for censored subjects in the active
treatment group is higher (more conservative) or similar to those
in the control group. In superiority trials, the control-based models
are appealing to clinical scientists since they would procedure conservative
conclusions about the treatment effect if the experimental treatment
is hypothesized to be better than the control treatment. }%
\begin{comment}
Atkinson 2019

This is particularly the case when censoring occurs because patients
change, or revert, to the usual (ie, reference) standard of care.
Recent work has shown how such questions can be addressed for trials
with continuous outcome data and longitudinal follow-up, using reference-based
multiple imputation.

Reference-based imputation has two advantages: (a) it avoids the user
specifying numerous parameters describing the distribution of patients'
postwithdrawal data and (b) it is, to a good approximation, information
anchored, so that the proportion of information lost due to missing
data under the primary analysis is held constant across the sensitivity
analyses.
\end{comment}

Another important question arises regarding the estimand of interest
for treatment comparison in the presence of missing data. Following
the International Council for Harmonization (ICH) E9 (R1) addendum,
estimands should be clearly defined which describe the quantity to
be estimated including how to handle intercurrent events such as premature
dropout \citep{ICHE9R1}. In this article, we \textcolor{black}{consider
a }\textit{\textcolor{black}{de facto}}\textcolor{black}{{} or treatment
policy strategy}, which evaluates treatment effect for all randomized
patients on time to event endpoint regardless of the deviation of
treatment such as taking rescue medication or treatment switch. When
time to event data are censored due to premature dropout, the primary
analysis often assumes CAR.  

For survival sensitivity analysis using $\delta$-adjusted models,
\citet{lipkovich2016sensitivity} considered a marginal  proportional
hazards parameter, an additional structural assumption entailing a
constant ratio of the hazard rates between the treatment groups. However,
this parameter may be misleading \citep{hernan2010hazards} if the
proportional hazards assumption is violated as in the $\delta$-adjusted
models. Alternatively, we consider a broad class of treatment effect
estimands defined as functionals of the survival functions, such as
the \textit{restricted mean survival time} (RMST, \citealp{chen2001causal}), %royston2013restricted,zhao2016restricted}),
i.e., the expectation of survival time restricted to a finite time
$\tau$. Instead of focusing on a constant hazards ratio, the RMST
provides a time-evolving profile of survival times for evaluating
the treatment effect, without requiring additional model assumptions.

To implement sensitivity analysis, multiple imputation (MI, \citealp{rubin1987multiple})
is the most popular method. It consists of three steps: first, fill
the missing values by plausible values to create multiple complete
datasets; second, apply standard full-sample methods to analyze the
multiple imputed datasets; and third, use Rubin's combining rule to
summarize the results for inference. Because of its intuitive appeal,
MI is recommended by the NRC as one of its preferred approaches of
addressing missing data \citep{little2012prevention}. Indeed, MI
provides a valuable tool to handle missing data arising from clinical
trials; however, a major challenge arises for inferences. Many studies
have realized that Rubin's variance estimator is not always consistent
for general purposes (e.g., \citealp{yang2016mi}). A sufficient condition
for the validity of the MI inference is the congeniality condition
\citep{meng1994multiple}. Roughly speaking, it requires the imputation
model to be correctly specified and the subsequent analysis to be
compatible with the imputation model. Even with a correctly specified
imputation model, \citet{yang2016mi} showed that MI is not necessarily
congenial for the method of moments estimation, so some common statistical
procedures may be incompatible with MI. This phenomenon becomes pronounced
for adopting MI for general sensitivity analysis in clinical trials.

\citet{lu2015comparison} and \citet{liu2016analysis} demonstrated
that Rubin's combining rule is often conservative in control-based
imputation. To overcome the conservative of Rubin's combining rule,
several authors suggested the non-parametric bootstrap to obtain the
standard errors \citep{lu2015comparison};
however, the non-parametric bootstrap requires repeating imputation
and analysis for all bootstrap samples and therefore causes huge computation
burden. Recently, \citet{guan2019unified} proposed the \textcolor{black}{wild-bootstrap}
inference of a martingale representation of the MI estimator; however,
their method is only applicable to continuous or binary outcomes but
not censored survival outcomes. \textcolor{black}{The standard nonparametric
bootstrap requires resampling individual observations and repeating
the imputation and analysis procedures; while the wild-bootstrap uses
an auxiliary zero-mean, unit variance random multiplier on the martingale
residuals for variance estimation. }

In this article, we propose a unified framework of survival sensitivity
analysis via MI. Specifically, the missing event times are imputed
by a $\delta$-adjusted or control-based Cox model for each treatment
group. We derive a novel martingale representation of the proposed
MI estimator. The martingale  representation is inspired by the sequential
construction of the MI estimator, namely, model parameter estimation
and imputations. This new representation invokes the easy-to-implement
wild-bootstrap inference. In contrast to Rubin's combining rule, the
wild-bootstrap inference has a theoretical guarantee for consistency.
Moreover, unlike the non-parametric bootstrap, we do not require repeating
imputation and analysis for the bootstrap resamples and therefore
largely reduce the computation burden. The new SMIM (\textit{S}urvival
sensitive analysis using \textit{M}ultiple \textit{I}mputation and
\textit{M}artingale) framework is fairly flexible to accommodate a
wide collection of censoring assumptions and treatment effect estimands.

The rest of this paper proceeds as follows. Section \ref{sec:Basic-Setup}
introduces notation, estimands, MI, and an outline of the proposed
SMIM framework. Section \ref{sec:D-CBimpM} presents sensitivity analysis
using the $\delta$-adjusted and control-based Cox models via MI.
Section \ref{sec:Wild-Bootstrap-Inference} derives the martingale
representation of the MI estimator and the wild bootstrap inference.
Section \ref{sec:Applications} applies the novel estimator to two
clinical trials. Section \ref{sec:Concluding-Remarks} concludes.
The Web Appendix contains the proofs\textcolor{black}{{} and simulation
studies}. An open source R package $\mathtt{smim}$ is available at
\url{https://github.com/elong0527/smim}.

\section{Setup \label{sec:Basic-Setup}}

\subsection{Notation and estimands}

Without loss of generality, we focus on randomized clinical trials
that compare a new treatment to a control treatment. We assume that
the subjects constitute a random sample from a larger population.
Let $X_{i}$ be a vector of covariates for subject $i$, and let $A_{i}$
be a binary treatment, $1$ for the active treatment and $0$ for
the control treatment. Let $T_{i}$ and $C_{i}$ denote the time to
a clinical event and the time to censoring, respectively. The full
set of variables is $F_{i}=(X_{i},A_{i},T_{i},C_{i})$. In the presence
of censoring, denote $U_{i}=T_{i}\wedge C_{i}$, where $\wedge$ represents
the minimal of two values, and $I_{i}=\bone(T_{i}\leq C_{i})$, where
$\bone(\cdot)$ is the indicator function taking value $1$ if its
argument is true and $0$ otherwise. To distinguish different reasons
for censoring, denote $R_{i}=1$ if censoring is due to administrative
reasons and $R_{i}=2$ if censoring is due to premature dropout. Extension
to more than two reasons is straightforward at the expense of heavier
notation. The observed set of variables is $O_{i}=\{X_{i},A_{i},U_{i},I_{i},(1-I_{i})R_{i}\}$.
We use $O_{1:k}$ to denote the $k$ copies $\{O_{1},\ldots,O_{k}\}$.
For the total of $n$ subjects, let $n_{1}=\sum_{i=1}^{n}A_{i}$ and
$n_{0}=\sum_{i=1}^{n}(1-A_{i})$. For notational convenience, let
the treated subjects be indexed by $i=1,\ldots,n_{1}$, and let the
control subjects be indexed by $i=n_{1}+1,\ldots,n.$

For treatment comparison, \textcolor{black}{define $\lambda_{a}(t)=\lim_{h\rightarrow0}h^{-1}\bP\left(t\leq T<t+h\mid T\geq t,A=a\right)$
and }$S_{a}(t)=\bP(T\geq t\mid A=a)$\textcolor{black}{{} as the treatment-specific
hazard rate }and survival function\textcolor{black}{{} at time $t$,
respectively, }for $a=0,1$. Under a proportional hazards assumption
\citep{hernan2000marginal}, one can focus on\textcolor{black}{{} estimating
log hazard ratio $\beta=\log\{\lambda_{1}(t)/\lambda_{0}(t)\}$}.
However, the proportional hazards assumption may be problematic, especially
when two survival curves cross. In particular, in sensitivity analysis,
the hazard ratios are constructed to be different before and after
patient dropout and hence the proportional hazards assumption is violated.
In this case, $\beta$ represents the overall average of the log hazard
ratios over a certain time period, which varies as the time period
changes \citep{hernan2010hazards}. Thus, $\beta$ lacks a clear interpretation.

Alternatively, we focus on treatment effect estimands defined as functionals
of treatment-specific survival distributions. Denote such a functional
as $\Delta_{\tau}=\Psi_{\tau}\{S_{1}(t),S_{0}(t)\},$ which may depend
on some pre-specified constant $\tau$. This formulation covers a
broad class of estimands favored in the context of non-proportional
hazards\textcolor{black}{; see examples of $\Delta_{\tau}$ below.}

\begin{example}[Treatment effect estimands]\label{eg}With a proper
choice of $\Psi_{\tau}(\cdot)$, $\Delta_{\tau}$ represents the following
measures of treatment effect: 
\begin{enumerate}
\item the difference in survival at a fixed time point $\tau$, $\Delta_{\tau}=S_{1}(\tau)-S_{0}(\tau)$; 
\item the difference of treatment-specific $\tau$-RMSTs (restrictive mean
survival times) $\Delta_{\tau}=\mu_{1,\tau}-\mu_{0,\tau},$ where
$\mu_{a,\tau}=\int_{0}^{\tau}S_{a}(t)\de t$ for $a=0,1$; 
\item the difference of weighted $\tau$-RMSTs $\Delta_{\tau}=\int_{0}^{\tau}\omega(t)\{S_{1}(t)-S_{0}(t)\}\de t$,
where the non-negative weight function $\omega(t)$ provides differentiable
importance at different times; 
\item the ratio of $\tau$-RMTLs (restrictive mean time lost) $\Delta_{\tau}=\{\tau-\int_{0}^{\tau}S_{1}(t)\de t\}/\{\tau-\int_{0}^{\tau}S_{0}(t)\de t\}$; 
\item the difference of $\tau$th quantiles (e.g., medians) of survivals
$\Delta_{\tau}=q_{1,\tau}-q_{0,\tau}$, where $q_{a,\tau}=\inf_{q}\{S_{a}(q)\leq\tau\}$. 
\end{enumerate}
\end{example}

\textcolor{black}{For identifiability, $\tau$ should be chosen properly.
For the estimands in }\textit{\textcolor{black}{a)--d)}}\textcolor{black}{,
we restrict $\tau$ to be smaller than  the minimum of
the largest observed survival times in the two treatment groups, say
$t_{\min}$, because the observed data can not provide information
about both treatment-specific survival distributions beyond this cut-off
value. Similarly, for the $\tau$th quantiles in }\textit{\textcolor{black}{e)}}\textcolor{black}{,
we require $\tau>$$\max\{S_{0}(t_{\min}),S_{1}(t_{\min})\}$. }

\subsection{Simple full-sample estimator and asymptotic linearity }

If the event times are fully observed, standard full-sample estimators
can apply. To estimate $S_{a}(t)$, a simple estimator is the sample
proportion $\hat{S}_{a,n}(t)=n_{a}^{-1}\sum_{i=1}^{n}\bone(A_{i}=a)\bone(T_{i}\geq t)$,
for $a=0,1$. Then, a plug-in estimator of $\Delta_{\tau}$ is $\hat{\Delta}_{\tau,n}=\Psi_{\tau}\{\hat{S}_{1,n}(t),\hat{S}_{0,n}(t)\}$.

To establish a unified framework, it is important to note that $\hat{\Delta}_{\tau,n}$
is asymptotically linear for all estimands given in Example \ref{eg}.
Under mild regularity conditions, we have
\begin{equation}
\hat{\Delta}_{\tau,n}-\Delta_{\tau}=\sum_{a=0}^{1}\int_{0}^{\tau}\psi_{a}(t)\left\{ \hat{S}_{a,n}(t)-S_{a}(t)\right\} \de t+o_{p}(n^{-1/2}),\label{eq:linear form}
\end{equation}
for bounded variation functions $\psi_{a}(\cdot)$.

\begin{lemma}[Asymptotic linear characterizations]

For all estimands in Example \ref{eg}, the full-sample estimators
have the following asymptotic linear characterizations. 
\begin{enumerate}
\item For the difference in the survivals at a fixed time point $\tau$,
$\hat{\Delta}_{\tau,n}=\hat{S}_{1,n}(\tau)-\hat{S}_{0,n}(\tau)$,
corresponding to (\ref{eq:linear form}) with $\psi_{1}(t)=-\psi_{0}(t)=\bone(t=\tau)$. 
\item For the difference of the treatment-specific $\tau$-RMSTs, $\hat{\Delta}_{\tau,n}=\int_{0}^{\tau}\left\{ \hat{S}_{1,n}(t)-\hat{S}_{0,n}(t)\right\} \de t$,
corresponding to (\ref{eq:linear form}) with $\psi_{1}(t)=-\psi_{0}(t)=1$. 
\item For the difference of weighted $\tau$-RMSTs, $\hat{\Delta}_{\tau,n}=\int_{0}^{\tau}\omega(t)\left\{ \hat{S}_{1,n}(t)-\hat{S}_{0,n}(t)\right\} \de t$,
corresponding to (\ref{eq:linear form}) with $\psi_{1}(t)=-\psi_{0}(t)=\omega(t)$. 
\item \label{enu:ratio}For the ratio of $\tau$-RMTLs, $\hat{\Delta}_{\tau,n}=\left\{ \tau-\int_{0}^{\tau}\hat{S}_{1,n}(t)\de t\right\} /\left\{ \tau-\int_{0}^{\tau}\hat{S}_{0,n}(t)\de t\right\} $,
corresponding to (\ref{eq:linear form}) with $\psi_{1}(t)=-\{\tau-\int_{0}^{\tau}S_{0,n}(u)\de u\}^{-1}$
and $\psi_{0}(t)=-\Delta_{\tau}\{\tau-\int_{0}^{\tau}S_{0,n}(u)\de u\}^{-1}$. 
\item \label{enu:quantile}For $\Delta_{\tau}=q_{1,\tau}-q_{0,\tau}$, $\hat{\Delta}_{\tau,n}=\hat{q}_{1,\tau}-\hat{q}_{0,\tau}$,
where $\hat{q}_{a,\tau}=\inf_{q}\{\hat{S}_{a,n}(q)\leq\tau\}$, corresponding
to (\ref{eq:linear form}) with $\psi_{1}(t)=\{\dot{S}_{1}(q_{1,\tau})\}^{-1}\bone(t=q_{1,\tau})$
and $\psi_{0}(t)=-\{\dot{S}_{0}(q_{0,\tau})\}^{-1}\bone(t=q_{0,\tau})$,
where $\dot{S}_{a}(q)=\de S_{a}(q)/\de q$. 
\end{enumerate}
\end{lemma}

For the ratio-type estimator in \textit{d)}, the asymptotic linear
characterization can be obtained by the Taylor expansion. For the
quantiles in \textit{e)}, under certain regularity conditions \citep[e.g.,][]{francisco1991quantile},
we can express $\hat{q}_{a,\tau}$ as 
\begin{equation}
\hat{q}_{a,\tau}-q_{a,\tau}=\frac{\hat{S}_{a,n}(q_{a,\tau})-S_{a}(q_{a,\tau})}{\dot{S}_{a}(q_{a,\tau})}+o_{P}(n^{-1/2}).\label{eq:quantile nni}
\end{equation}
Expression (\ref{eq:quantile nni}) is called the Bahadur-type representation
for $\hat{q}_{a,\tau}$. Then, the asymptotic linear characterization
in \textit{e)} follows.

\subsection{MI}

To facilitate applying full-sample estimators, MI creates multiple
complete datasets by filling in missing values. MI proceeds as follows. 
\begin{description}
\item [{Step$\ $MI-1.}] Create $m$ complete datasets by filling in missing
times to event with imputed values generated from an imputation model.
Specifically, to create the $j$th imputed dataset, generate $T_{i}^{*(j)}$
from the imputation model for each missing $T_{i}$. Further discussions
on the imputation models are provided in Section \ref{sec:D-CBimpM}. 
\item [{Step$\ $MI-2.}] Apply a full-sample estimator of $\text{\ensuremath{\Delta_{\tau}}}$
to each imputed dataset. Denote the point estimator applied to the
$j$th imputed dataset by $\text{\ensuremath{\hat{\Delta}_{\tau}^{(j)}}}$,
and the variance estimator by $\hat{V}^{(j)}$. 
\item [{Step$\ $MI-3.}] Use Rubin's combining rule to summarize the results
from the multiple imputed datasets. The MI estimator of $\Delta_{\tau}$
is $\text{\ensuremath{\hat{\Delta}_{\tau,\mi}}}=m^{-1}\sum_{j=1}^{m}\hat{\Delta}_{\tau}^{(j)}$,
and Rubin's variance estimator is 
\begin{equation}
\hat{V}_{\mi}(\ensuremath{\hat{\Delta}_{\tau,\mi}})=\frac{m+1}{(m-1)m}\sum_{j=1}^{m}(\hat{\Delta}_{\tau}^{(j)}-\ensuremath{\hat{\Delta}_{\tau,\mi}})^{2}+\frac{1}{m}\sum_{j=1}^{m}\hat{V}^{(j)}.\label{eq:rubin-1}
\end{equation}
\end{description}
It is well known that Rubin's combining rule may overestimate the
variance of the MI estimator when the full-sample estimators are not
self-efficient. To see the problem, consider the following decomposition
$\hat{\Delta}_{\tau,\mi}-\ensuremath{\Delta_{\tau}}=(\hat{\Delta}_{\tau,\mi}-\hat{\Delta}_{\tau,n})+(\hat{\Delta}_{\tau,n}-\ensuremath{\Delta_{\tau})},$
and therefore the variance of $\hat{\Delta}_{\tau,\mi}$ is 
\[
\V(\hat{\Delta}_{\tau,\mi})=\V\left(\hat{\Delta}_{\tau,\mi}-\hat{\Delta}_{\tau,n}\right)+\V\left(\hat{\Delta}_{\tau,n}\right)+2\cov\left(\hat{\Delta}_{\tau,\mi}-\hat{\Delta}_{\tau,n},\hat{\Delta}_{\tau,n}\right).
\]
The two terms in Rubin's variance estimator (\ref{eq:rubin-1}) estimate
$\V(\hat{\Delta}_{\tau,\mi}-\hat{\Delta}_{\tau,n})$ and $\V(\hat{\Delta}_{\tau,n})$,
respectively. It presumes the covariance term is zero, which, however,
is not true in general. In this case, Rubin's combining rule is not
consistent. \citet{lu2015comparison} and \citet{liu2016analysis}
demonstrated this issue in the sensitivity analysis using control-based
imputation.

We provide an alternative decomposition of the MI estimator, which
invokes the wild bootstrap for consistent variance estimation for
general imputation models and estimands. Before we delve into the
technical details, we provide an outline of the proposed SMIM framework
below. 

\subsection{Outline of the proposed SMIM framework }

In Step MI-1, we consider a flexible class of $\delta$-adjusted and
control-based Cox imputation models for sensitivity analysis. For
example, the $\delta$-adjusted Cox model assumes \textcolor{black}{the
treatment-specific hazard rate of failing at time $t$ }is $\lambda_{a}(t\mid X_{i})$
without premature dropout and $\delta\lambda_{a}(t\mid X_{i})$ after
dropout, for $a=0,1$. Importantly, under the $\delta$-adjusted Cox
model, we do not impose the restrictive proportional hazards assumption
on the treatment effect. More details will be provided in Section
\ref{sec:D-CBimpM}.

Based on the MI with Rubin's combining rule in Step MI-3, the variance
estimator overestimates the true variance of $\hat{\Delta}_{\tau,\mi}$.
For rectification, we propose a wild bootstrap variance estimator
\citep{wu1986jackknife,liu1988bootstrap} to replace Rubin's combining
rule; Theorem \ref{th:WBconsistency} in Section \ref{sec:Wild-Bootstrap-Inference}
shows that the proposed variance estimator is consistent for general
imputation models and treatment effect estimands. The consistency
ensures the confidence intervals have proper coverage properties.
The wild bootstrap procedure does not require \textcolor{black}{repeating}
the missing data imputation step (i.e., Step MI-1) and \textcolor{black}{recalculating}
the point estimator (i.e., Step MI-2) using resampling data, therefore
it is computationally efficient compared with the naive bootstrap.

The wild bootstrap variance estimator is motivated by a novel martingale
representation of the MI estimator. Specifically, we show in Section
\ref{sec:D-CBimpM} that the MI estimator of $\Delta_{\tau}$ can
be represented as 
\[
n^{1/2}(\text{\ensuremath{\hat{\Delta}_{\tau,\mi}}}-\Delta_{\tau})=\sum_{k=1}^{(1+m)n}\xi_{n,k}+o_{p}(1),
\]
where the series $\{\sum_{i=1}^{k}\xi_{n,i},\;1\le k\le(1+m)n\}$
along with properly defined $\sigma$-fields is a martingale array.
This representation invokes the wild bootstrap procedure that provides
valid variance estimation and inference of the MI estimator of $\Delta_{\tau}$
\citep{pauly2011weighted}. 

\section{Delta-adjusted and control-based models\label{sec:D-CBimpM}}

\subsection{Primary analysis with the CAR benchmark assumption }

To motivate the imputation models for sensitivity analysis, we first
consider the CAR assumption that $C_{i}\perp\!\!\!\perp T_{i}\mid(A_{i},X_{i})$.
Under CAR, we have 
\begin{eqnarray*}
\lambda_{a}(t\mid X_{i}) & = & \lim_{h\rightarrow0}h^{-1}\bP\left(t\leq T_{i}<t+h\mid T_{i}\geq t,X_{i},A_{i}=a\right)\\
 & = & \lim_{h\rightarrow0}h^{-1}\bP\left(t\leq U_{i}<t+h,I_{i}=1\mid U_{i}\geq t,X_{i},A_{i}=a\right),
\end{eqnarray*}
for $a=0,1$. From $\lambda_{a}(t\mid X_{i})$, we can derive the
survival function for the subject $i$ as $S_{a}(t\mid X_{i})=\exp\{-\int_{0}^{t}\lambda_{a}(u\mid X_{i})\de u\}$.
For regularity, we impose a positivity condition for $S_{a}(t\mid X_{i})$.

\begin{assumption}[Positivity]\label{assump:pos}

There exists a constant $c$ such that with probability one, $S_{a}(t\mid X_{i})\geq c>0$
for $t$ in $[0,\tau]$ and $a=0,1$.

\end{assumption}

Following most of the survival analysis literature \citep[e.g.,][]{chen2001causal},
we posit a conditional treatment-specific Cox regression with covariate
$X_{i}$; i.e., 
\begin{equation}
\lambda_{a}(t\mid X_{i})=\lambda_{a}(t)e^{\beta_{a}^{\T}X_{i}},\label{eq:model-1}
\end{equation}
where $\lambda_{a}(t)$ is an unknown baseline hazard function and
$\beta_{a}$ is a vector of unknown parameters for $a=0,1$. Importantly,
under model (\ref{eq:model-1}), we do not impose the restrictive
proportional hazards assumption on the treatment effect because both
$\lambda_{a}(t)$ and $\beta_{a}$ can be different for the two treatment
groups. %Other flexible survival models can also be considered \citep[e.g.,][]{zeng2007maximum,yin2008partially}.
Let $\theta=\{\lambda_{a}(\cdot),\beta_{a}:a=0,1\}$ summarize the
infinite-dimensional parameter in the Cox model. Under CAR, we can
estimate $\theta$ from the standard software such as ``coxph''
in R. %(R Development Core Team, 2012) \nocite{R:2010}.

We adopt the counting process framework \citep{andersen1982cox} to
introduce the estimators and their large sample properties. Define
the counting process $N_{i}(t)=\bone(U_{i}\leq t,I_{i}=1)$ of observing
the event and the at-risk process $Y_{i}(t)=\bone(U_{i}\geq t)$.
Let $\hat{\beta}_{a}$ be the maximum partial likelihood estimator
of $\beta_{a}$, for $a=0,1$. We can estimate the cumulative baseline
hazard, $\Lambda_{a}(t)=\int_{0}^{t}\lambda_{a}(u)\de u$ by the \citet{breslow1974covariance}
estimator 
\[
\hat{\Lambda}_{a}(t)=\int_{0}^{t}\hat{\lambda}_{a}(u)\de u,\ {\color{black}\hat{\lambda}_{a}(u)\de u=\frac{\sum_{j=1}^{n}\bone(A_{j}=a)\de N_{j}(u)}{\sum_{j=1}^{n}\bone(A_{j}=a)e^{\hat{\beta}_{a}^{\T}X_{j}}Y_{j}(u)}},
\]
and estimate $S_{a}(t\mid X_{i})$ by $\hat{S}_{a}(t\mid X_{i})=\exp\left\{ -\hat{\Lambda}_{a}(t)e^{\hat{\beta}_{a}^{\T}X_{i}}\right\} .$
Under standard regularity conditions, $n^{1/2}\{\hat{S}_{a}(t\mid X_{i})-S_{a}(t\mid X_{i})\}$
converges uniformly to a Gaussian process in $[0,\tau]$; see, e.g.,
\citet{andersen1982cox}.

\textcolor{black}{The CAR assumption is not testable and may be questionable
for censoring due to premature dropout. Sensitivity analysis is critical
to assess the robustness of study conclusions to CAR. }

\subsection{Sensitivity analyses with $\delta$-adjusted and control-based models}

Toward that end, we propose sensitive analysis using a wide range
of imputation models including the $\delta$-adjusted models and the
control-based models.

\begin{assumption}[Delta-adjusted Cox model]\label{assumption:Delta-adjusted-model}

T\textcolor{black}{he treatment-specific hazard rate of failing at
time $t$ }is $\lambda_{a}(t\mid X_{i})$ given in (\ref{eq:model-1})
without premature dropout and is $\delta\lambda_{a}(t\mid X_{i})$
after premature dropout ($R_{i}=2$), for $a=0,1$, where $\delta>0$.

\end{assumption}

It can be seen that $\delta$ quantifies the degree of the departure
from the CAR assumption. If $\delta=1$, we have CAR. If $\delta>1$,
the hazard increases after dropout, indicating a worsening of condition
after dropout. If $\delta<1$, the hazard decreases after dropout,
indicating an improvement of condition after dropout. The larger magnitude
of $\delta$, the larger deviation from CAR. Without retrieving information
for the non-administratively censored subjects, $\delta$ can not
be ascertained. Therefore, it is recommended to vary $\delta$ in
a wide plausible range of values for sensitivity analysis. To fix
ideas, we use the same $\delta$ for both treatment groups, but it
is easy to accommodate different $\delta$ values depending on the
worsening/improvement condition for different treatment groups. For
example, if the control group is a placebo group, it is reasonable
to choose $\delta$ to be one for the control subjects who was non-administratively
censored. We illustrate the use of different $\delta$ for different
treatment groups in Sections \ref{sec:sim} and \ref{sec:Applications}. %sy:check 

Control-based models \citep[e.g.,][]{carpenter2013analysis} are another
popular class of sensitivity models. These models are appealing because
of their reduced bias in favor of the experimental treatment. 

\begin{assumption}[Control-based Cox model]\label{assumption:control Cox model}

T\textcolor{black}{he treatment-specific hazard rate of failing at
time $t$ }is $\lambda_{a}(t\mid X_{i})$ given in (\ref{eq:model-1})
for $a=0,1$ and is $\delta\lambda_{0}(t\mid X_{i})$ after dropout
($R_{i}=2$) for the treated, where $\delta\leq1$. 

\end{assumption}

\textcolor{black}{The control-based Cox model with $\delta=1$ becomes
the jump-to-reference model \citep{atkinson2019reference}. It assumes
that censored subjects on the active arm follow the same distribution
as similar subjects in the control group after the censored time.
This model is, for example, plausible for superiority trials if subjects
on the control arm received the standard care and censoring on the
active arm is because subjects revert to the standard of care. For
generality, we also allow $\delta$ to be less than one, such that
the treatment effect can be bracketed by the treatment effect under
CAR and that for the control arm \citep{lu2015comparison}. }

In fact, \textcolor{black}{censoring due to dropout can be interpreted
as a time-dependent binary covariate, }\textcolor{black}{and} $\delta$-adjusted
and control-based sensitivity models entail \textcolor{black}{time-dependent
Cox models}. Let the history of the information up to time $t$ be
$H_{i}(t)=\{X_{i},R_{i},N_{i}(u),Y_{i}(u):u<t\}$. Because we use
$R_{i}=2$ to indicate premature dropout, Assumption \ref{assumption:Delta-adjusted-model}
describes the time-dependent Cox model with the hazard function 
\begin{equation}
\lambda_{1}\{t\mid H_{i}(t);\delta,\theta\}=\lambda_{1}(t)\delta^{\bone(R_{i}=2\ \&\ t>U_{i})}e^{\beta_{1}^{\T}X_{i}}.\label{eq:delta-cox}
\end{equation}
Assumption \ref{assumption:control Cox model}
describes the time-dependent Cox model with the hazard function, for $a=0,1$, 
\begin{equation}
\lambda_{a}\{t\mid H_{i}(t);\delta,\theta\}=\begin{cases}
\lambda_{0}(t)e^{\beta_{0}^{\T}X_{i}} & \text{if }a=0,\\
\delta\lambda_{0}(t)e^{\beta_{0}^{\T}X_{i}} & \text{if }a=1,R_{i}=2,t>U_{i}\\
\lambda_{1}(t)e^{\beta_{1}^{\T}X_{i}} & \text{otherwise}.
\end{cases},\label{eq:cb-cox}
\end{equation}

The \textit{de facto} estimand for treatment policy takes into account
the likely attenuation of the treatment effect after dropout. By (\ref{eq:delta-cox})
and (\ref{eq:cb-cox}), the \textit{de facto} survival function is
\[
S_{a}^{\sen}(t)=\E\left[\exp\left\{ -\int_{0}^{t}\lambda\{u\mid H_{i}(u);\delta,\theta\}\de u\right\} \right],
\]
for $a=0,1$. Here we use the superscript ``sen'' to denote either
``$\delta$-adj'' or ``cb'' for the delta-adjusted or control-based
sensitivity model. The \textit{de facto} treatment effect estimand
becomes $\Delta_{\tau}^{\sen}=\Psi_{\tau}\{S_{1}^{\sen}(t),S_{0}^{\sen}(t)\}$.
If the sensitivity parameter $\delta$ is not one, $\Delta_{\tau}^{\sen}$
differs from $\Delta_{\tau}$ in general. By varying $\delta$ over
a certain range, $\Delta_{\tau}^{\sen}$ provides valuable insights
on the impact of possible departures from CAR, allowing an investigator
to assess the extent to which the censoring assumption alters the
treatment effect estimator.

MI requires generating the missing values from the imputation model
in Step MI-1. From (\ref{eq:delta-cox}) or (\ref{eq:cb-cox}), one
can derive the conditional survival function $S_{a}\{t\mid H_{i}(t);\delta,\theta\}$
for imputation. Consider the $\delta$-adjusted model for example,
if a treated subject $i$ withdrew from the treatment, the conditional
survival at $t>U_{i}$ is 
\begin{equation}
S_{1}\{t\mid H_{i}(t);\delta,\theta\}=e^{-\int_{U_{i}}^{t}\delta\lambda_{1}(u\mid X_{i})\de u}.\label{eq:model-2}
\end{equation}
Unlike the parametric models, sampling from the semiparametric Cox
model is difficult. Following \citet{lipkovich2016sensitivity}, we
introduce a general inverse transform sampling scheme. Suppose we
would like to generate $T_{i}^{*}$ from (\ref{eq:model-2}) for $t\geq U_{i}$.
First, generate a random number $u_{i}$ from Unif$[0,p_{i}]$, where
$p_{i}=\{S_{1}(U_{i}\mid X_{i})\}^{\delta}$. Second, solve $\{S_{1}(T_{i}^{*}\mid X_{i})\}^{\delta}=u_{i}$
for $T_{i}^{*}$. Then, we show that given the observed data $O_{1:n}$,
\begin{multline*}
\bP\left(T_{i}^{*}\geq t\mid O_{1:n}\right)=\bP\left[\{S_{1}(T_{i}^{*}\mid X_{i})\}^{\delta}\leq\{S_{1}(t\mid X_{i})\}^{\delta}\mid O_{1:n}\right]\\
=\bP\left[u_{i}\leq\{S_{1}(t\mid X_{i})\}^{\delta}\mid O_{1:n}\right]=\{S_{1}(t\mid X_{i})\}^{\delta}/p_{i}=e^{-\int_{U_{i}}^{t}\delta\lambda_{1}(u\mid X_{i})\de u}
\end{multline*}
is the target imputation model (\ref{eq:model-2}). 

In practice, we need numerical approximations to obtain $T_{i}^{*}$.
Let $T_{a,\max}$ be the largest observed event time in treatment
group $a$ for $a=0,1$. Because $S_{a}(t\mid X_{i})$ is semiparametric,
$\hat{S}_{a}(t\mid X_{i})$ is only available for $t\leq T_{a,\max}$.
Thus we require $\tau$ to be smaller than $\tilde{T}_{\max}=T_{0,\max}\wedge T_{1,\max}$,
and then the imputed value $T_{i}^{*}$ can be truncated at $\tilde{T}_{\max}$. 

To summarize, the MI procedure for $\delta$-adjusted and control-based
imputations proceeds as follows. 
\begin{description}
\item [{Step$\ $MI-1-1.}] Fit a Cox model assuming CAR; denoted by $S_{a}(t\mid X_{i};\hat{\theta})$. 
\item [{Step$\ $MI-1-2.}] For administratively censored subject $i$ with
$(A_{i},I_{i},R_{i})=(a,0,1)$, compute $p_{i}=S{}_{a}(U_{i}\mid X_{i};\hat{\theta})$.
Draw a uniform random value $u_{i}\sim$ Unif$[0,p_{i}]$. Impute
the event time $T_{i}^{*}$ as the solution of $u_{i}=S_{a}(t\mid X_{i};\hat{\theta})$.
Numerically, we use $T_{i}^{*}=\arg\max_{t\in\mathcal{T}}\{S_{a}(t\mid X_{i};\hat{\theta})\geq u_{i}\}$,
where $\mathcal{T}$ is the set of realized times to event or censoring
with the largest value being $\tilde{T}_{\max}$. This will ensure
that the imputed event time falls between the censoring time and $\tilde{T}_{\max}$. 
\end{description}
For $\delta$-adjusted imputation model, Step MI-1-3 proceeds as follows. 
\begin{description}
\item [{Step$\ $MI-1-3.}] For non-administratively censored subject $i$
with $(A_{i},I_{i},R_{i})=(a,0,2)$, compute $p_{i}=\{S_{a}(U_{i}\mid X_{i};\hat{\theta})\}^{\delta}$.
Draw a uniform random value $u_{i}\sim$ Unif$[0,p_{i}]$. Impute
the event time $T_{i}^{*}$ as the solution of $u_{i}=\{S_{a}(t\mid X_{i};\hat{\theta})\}^{\delta}$.
Numerically, we use $T_{i}^{*}=\arg\max_{t\in\mathcal{T}}[\{S_{a}(t\mid X_{i};\hat{\theta})\}^{\delta}\geq u_{i}]$. 
\end{description}
For control-based imputation model, Step MI-1-3 proceeds as follows. 
\begin{description}
\item [{Step$\ $MI-1-3'.}] For non-administratively censored subject $i$
with $(A_{i},I_{i},R_{i})=(0,0,2)$, draw $T_{i}^{*}$ by Step MI-1-3
with $a=0$ and $\delta=1$. For non-administratively censored subject
$i$ with $(A_{i},I_{i},R_{i})=(1,0,2)$, draw $T_{i}^{*}$ by Step
MI-1-3 with $a=0$ and $\delta$, i.e., using the corresponding distribution
in the control group. 
\end{description}

\section{Wild Bootstrap Inference based on Martingale Series\label{sec:Wild-Bootstrap-Inference}}

\subsection{A novel martingale representation}

For variance estimation, the key insight is that the MI estimator
is intrinsically created in a sequential manner: first, the imputation
model is fitted based on the observed data; second, the missing data
are drawn from the imputation model conditioned on the observed data.
This conceptualization leads to a martingale representation of the
MI estimator\textcolor{black}{{} by expressing the MI estimator in terms
of a series of random variables that have mean zero conditional on
the sigma algebra generated from the preceding variables. }We provide
heuristic steps below toward linearizing the MI estimator and forming
the proper sigma algebra and regulate details to the Web Appendix.

We first focus on treatment group $a=1$. To unify the notation, let
$T_{i}^{*(j)}$ denote the $j$th imputed value for subject $i$ if
subject $i$ was censored and the observed $T_{i}$ if we observe
subject $i$'s event time. By the imputation mechanism, $T_{i}^{*(j)}$
follows the conditional survival distribution $S_{1}\{t\mid H_{i}(t);\hat{\theta}\}$
for $t\geq U_{i}$, where $\theta=\{\lambda_{a}(\cdot),\beta_{a}:a=0,1\}$.
Then, for $t\in[0,\tau],$ it is insightful to express 
\begin{align}
 & n^{1/2}\left\{ \hat{S}_{1,\mi}(t)-S_{1}^{\sen}(t)\right\} =\frac{n^{1/2}}{mn_{1}}\sum_{j=1}^{m}\sum_{i=1}^{n}A_{i}\{\bone(T_{i}^{*(j)}\geq t)-S_{1}^{\sen}(t)\}\nonumber \\
 & \ \ \ \ =\frac{n^{1/2}}{mn_{1}}\sum_{j=1}^{m}\sum_{i=1}^{n}A_{i}\{1-Y_{i}(t)\}\left[\bone(T_{i}^{*(j)}\geq t)-S_{1}\{t\mid H_{i}(t);\hat{\theta}\}\right]\label{eq:MI}\\
 & \ \ \ \ +\frac{n^{1/2}}{n_{1}}\sum_{i=1}^{n}A_{i}\left[S_{1}\{t\mid H_{i}(t);\hat{\theta}\}-S_{1}^{\sen}(t)\right].\label{eq:MI2}
\end{align}
Here, we use the total sample size $n$ for scaling; we will use the
same scaling for the estimators for the control group and the treatment
effect.

We analyze the two terms in (\ref{eq:MI}) and (\ref{eq:MI2}), separately.
First, because the imputations are independent given the observed
data, it follows that the individual terms in (\ref{eq:MI}) are independent
mean-zero terms conditional on the observed data. Second, because
the term in (\ref{eq:MI2}) depends on $\hat{\theta}$, by exploiting
the counting process theory, we express 
\begin{eqnarray}
 &  & \frac{n^{1/2}}{n_{1}}\sum_{i=1}^{n}A_{i}\left[S_{1}\{t\mid H_{i}(t);\hat{\theta}\}-S_{1}^{\sen}(t)\right]\nonumber \\
 & = & \frac{n^{1/2}}{n_{1}}\sum_{i=1}^{n}A_{i}\left[Y_{i}(t)+\{1-Y_{i}(t)\}(1-I_{i})S_{1}\{t\mid H_{i}(t);\theta\}-S_{1}^{\sen}(t)\right]\label{eq:decomp-term2}\\
 &  & +\frac{n^{1/2}}{n_{1}}\sum_{i=1}^{n}A_{i}\phi_{11,i}(t)+\frac{n^{1/2}}{n_{1}}\sum_{i=1}^{n}(1-A_{i})\phi_{10,i}(t)+o_{p}(1),\label{eq:decomp}
\end{eqnarray}
where the exact expressions of $\phi_{11,i}(t)$ and $\phi_{10,i}(t)$
are given in Section \ref{sec:Proof-of-Thm1}. Importantly, $\phi_{11,i}(t)$
reflects the estimation of $\{\lambda_{1}(\cdot),\beta_{1}\}$, $\phi_{10,i}(t)$
reflects the estimation of $\{\lambda_{0}(\cdot),\beta_{0}\}$, and
$\E\{\phi_{11,i}(t)\}=\E\{\phi_{10,i}(t)\}=0$. Note that in the sensitivity
analysis using the $\delta$-adjusted models, the imputation for the
treated group uses the information only from the treated group, so
$\phi_{11,i}(t)\neq0$ and $\phi_{10,i}(t)=0$ for all $i$; while
in the sensitivity analysis using the control-based models, the imputation
for the treated group uses information from both treatment groups,
so $\phi_{11,i}(t)\neq0$ and $\phi_{10,i}(t)\neq0$ for all $i$.
Also, by definition, the expectation of the term in (\ref{eq:decomp-term2})
is zero. Together, $n^{1/2}\{\hat{S}_{1,\mi}(t)-S_{1}^{\sen}(t)\}$
decomposes into the summation of three terms (\ref{eq:MI}), (\ref{eq:decomp-term2}),
and (\ref{eq:decomp}) with (conditional) mean zero, and converges
to a Gaussian process in $[0,\tau]$. \textcolor{black}{Similarly,
we obtain a similar asymptotic linearization of $\hat{S}_{0,\mi}(t)$
given in }(\ref{eq:MI0-1})--(\ref{eq:MI0-3}).

We now leverage the unified linear characterization (\ref{eq:linear form})
to express the MI estimator for various treatment effect estimands.
Combining (\ref{eq:linear form}) and the above decompositions of
$\hat{S}_{1,\mi}(t)$ and $\hat{S}_{0,\mi}(t)$, we derive 
\begin{equation}
n^{1/2}(\hat{\Delta}_{\tau,\mi}-\Delta_{\tau})=n^{1/2}\left[\Psi_{\tau}\{\hat{S}_{1,\mi}(t),\hat{S}_{0,\mi}(t)\}-\Delta_{\tau}\right]=\sum_{k=1}^{(1+m)n}\xi_{n,k}+o_{p}(1),\label{eq:martingale}
\end{equation}
where
\begin{eqnarray}
\xi_{n,k} & = & \frac{n^{1/2}}{n_{1}}\int_{0}^{\tau}\psi_{1}(t)A_{i}\left[\phi_{11,i}(t)+Y_{i}(t)\vphantom{S_{1}^{\delta}}\right.+\left.\{1-Y_{i}(t)\}(1-I_{i})S_{1}\{t\mid H_{i}(t);\theta\}-S_{1}^{\sen}(t)\right]\de t,\nonumber \\
 &  & \text{for }k=i\ (1\leq i\leq n_{1}),\label{eq:xi-1}\\
\xi_{n,k} & = & \frac{n^{1/2}}{mn_{1}}\int_{0}^{\tau}\psi_{1}(t)A_{i}\{1-Y_{i}(t)\}[\bone(T_{i}^{*(j)}\geq t)-S_{1}\{t\mid H_{i}(t);\hat{\theta}\}]\de t,\nonumber \\
 &  & \text{for }k=n_{1}+(i-1)m+j\ (1\leq i\leq n_{1},1\leq j\leq m),\label{eq:xi-2}\\
\xi_{n,k} & = & \frac{n^{1/2}}{n_{0}}\int_{0}^{\tau}\psi_{0}(t)(1-A_{i})\left[\phi_{10,i}(t)+\phi_{0,i}(t)+Y_{i}(t)\vphantom{S_{0}^{\delta-adj}}\right.\label{eq:xi1}\\
 &  & +\left.\{1-Y_{i}(t)\}(1-I_{i})S_{0}\{t\mid H_{i}(t);\theta\}-S_{0}^{\sen}(t)\right]\de t,\nonumber \\
 &  & \text{for }k=(1+m)n_{1}+i,(n_{1}+1\leq i\leq n),\label{eq:xi-3}\\
\xi_{n,k} & = & \frac{n^{1/2}}{mn_{0}}\int_{0}^{\tau}\psi_{0}(t)(1-A_{i})\{1-Y_{i}(t)\}[\bone(T_{i}^{*(j)}\geq t)\ \ -S_{0}\{t\mid H_{i}(t);\hat{\theta}\}]\de t,\nonumber \\
 &  & \text{for }k=(1+m)n_{1}+n_{0}+(i-1)m+j\ (n_{1}+1\leq i\leq n,1\leq j\leq m).\label{eq:xi-4}
\end{eqnarray}
To gain intuitions, based on the decomposition in (\ref{eq:martingale}),
the first $n_{1}$ terms of $\xi_{n,k}$ contribute to the variability
of $\hat{S}_{1,\mi}$ because of the unknown parameters, and the next
$mn_{1}$ terms of $\xi_{n_{1},k}$ contribute to the variability
of $\hat{S}_{1,\mi}$ because of the imputations given the estimated
parameter values, reflecting the sequential MI procedure. Other terms
have similar explanations. 

We now form the proper sigma algebra $\{\F_{n,k}:1\leq k\leq(1+m)n\}$
such that $\E(\xi_{n,k}\mid\F_{n,k-1})=0$ for all $k$ and thus
\begin{equation}
\left\{ \sum_{i=1}^{k}\xi_{n,i},\F_{n,k},1\leq k\leq(1+m)n\right\} \ \text{is a martingale for each }n\geq1.\label{eq:martingale-1}
\end{equation}
We focus on the $\xi_{n,k}$ terms in (\ref{eq:xi-1}) and (\ref{eq:xi-2})
for treatment group $a=1$, because the discussion for the $\xi_{n,k}$
terms in (\ref{eq:xi-3}) and (\ref{eq:xi-4}) for treatment group
$a=0$ is similar and is presented in the Web Appendix. Obviously,
for $k=i$ $(1\leq i\leq n_{1})$ and $\xi_{n,k}$ in (\ref{eq:xi-1}),
we have $\E(\xi_{n,1})=0$ and $\E(\xi_{n,k}\mid O_{1:k-1})=\E(\xi_{n,k})=0$,
and thus we let $\F_{n,k}=\sigma(O_{1},\ldots,O_{k})$. For $k=n_{1}+(i-1)m+j$,
where $i=1,\ldots,n{}_{1}$ and $j=1,\ldots,m$, and $\xi_{n,k}$
in (\ref{eq:xi-2}), under the regularity conditions, we have $\E(\xi_{n,k}\mid O_{1},\ldots,O_{n_{1}},T_{1}^{*(1)},\ldots,T_{i}^{*(j)})=0$,
and thus we let $\F_{n,k}=\sigma(O_{1},\ldots,O_{n_{1}},T_{1}^{*(1)},\ldots,T_{i}^{*(j)})$. 

The martingale  representation allows us to characterize the asymptotic
distribution of $\hat{\Delta}_{\tau,\mi}$ with the proof presented
in Section \ref{sec:Proof-of-Thm1}.

\begin{theorem}\label{th:MI-point} Under Assumptions \ref{assump:pos},
\ref{assumption:Delta-adjusted-model}/\ref{assumption:control Cox model},
and \ref{asump:consistency} (regularity conditions), $n^{1/2}(\hat{\Delta}_{\tau,\mi}-\Delta_{\tau})\rightarrow\mathcal{N}(0,V_{\tau,\mi}^{\sen})$,
as $n\to\infty$, where $V_{\tau,\mi}^{\sen}$ is a finite variance
given in (\ref{eq:sigma}).

\end{theorem}

\subsection{Wild bootstrap for the MI estimator}

The martingale representation invokes the wild or weighted bootstrap
procedure \citep{wu1986jackknife,liu1988bootstrap} that provides
valid variance estimation and inference of the linear statistic for
martingale difference arrays. \citet{pauly2011weighted} proved the
validity of the wild bootstrap re-sampling under the conditions of
a general central limit theorem (CLT). \citet{guan2019unified} applied
the wild bootstrap for a martingale series in the context of causal
inference with observational studies.

Based on the martingale representation (\ref{eq:martingale}), we
propose the wild bootstrap procedure to estimate the variance of $\hat{\Delta}_{\tau,\mi}$.
The martingale representation relies on unknown quantities, requiring
approximations. We then estimate (i) $S_{a}^{\sen}(t)$ by $\hat{S}_{a,\mi}$,
(ii) $\phi_{11,i}(t)$, $\phi_{10,i}(t)$, and $\phi_{0,i}(t)$ by
$\hat{\phi}_{11,i}(t)$, $\hat{\phi}_{10,i}(t)$, and $\hat{\phi}_{0,i}(t)$,
and (iii) $S_{a}\{t\mid H_{i}(t);\theta\}$ by $S_{a}\{t\mid H_{i}(t);\hat{\theta}\}$,
for $a=0,1$.

Based on the above approximations, the wild bootstrap inference proceeds
as follows. 
\begin{description}
\item [{Step$\ $WB-1.}] Sample $u_{k}$, for $k=1,\ldots,(1+m)n$, that
satisfy $\E(u_{k}\mid O_{1:n})=0$, $\E(u_{k}^{2}\mid O_{1:n})=1$
and $\E(u_{k}^{4}\mid O_{1:n})<\infty$. 
\item [{Step$\ $WB-2.}] Compute the bootstrap replicate as $W_{L}^{*}=n^{-1/2}\sum_{k=1}^{(1+m)n}\hat{\xi}_{n,k}u_{k}$,
where $\hat{\xi}_{n,k}$ is the empirical version of $\xi_{n,k}$
by replacing the unknown quantities with their estimators and the
\textcolor{black}{one-dimensional integrals }by the numerical integration. 
\item [{Step$\ $WB-3.}] Repeat Steps 1 and 2 $B$ times, and estimate
the variance of $\hat{\Delta}_{\tau,\mi}$ by the sample variance
of these copies of $W_{L}^{*}$. 
\end{description}
\begin{remark}\label{rem:weight}

There are many choices for generating $\mu_{k}$, such as the the
standard normal distribution, Mammen's \citet{mammen1993bootstrap}
two point distribution, a simpler distribution with probability $0.5$
of being $1$ and probability $0.5$ of being $-1$, or the nonparametric
bootstrap weights. The wild bootstrap procedure is not sensitive to
the choice of the sampling distribution of $\mu_{k}$. We adopt the
standard normal distribution in the simulation study.

\end{remark}

\begin{remark} It is worth discussing the connection between the
martingale representation (\ref{eq:martingale}) and existing results
in the survival literature. Under CCAR, \citet{zhao2016restricted}
derived an asymptotic linearization for the RMST estimator and proposed
the perturbation-resampling variance estimation by adding independent
noises to the linearized terms. In this simpler case, by setting the
sensitivity parameter $\delta$ to be $1$ and omitting the imputation
step, our martingale representation with the first $n_{1}$ terms
reduces to their linearization. The slight difference lies in the
distribution for generating the resampling weights. In the wild bootstrap,
the resampling weight distribution has mean $1$; while in the perturbation,
the resampling weight distribution has mean $0$. \textcolor{black}{The
difference would only affect the center of the bootstrap replicates
of $\hat{\Delta}_{\tau,\mi}$ but not the variability and thus variance
estimation.} Our framework allows for CAR and sensitivity analysis
using $\delta$-adjustment/control-based models, taking into account
variability from both parameter estimation and imputation.

\end{remark}

Theorem \ref{th:WBconsistency} shows the asymptotic validity of the
above bootstrap inference method.

\begin{theorem}\label{th:WBconsistency} Under Assumptions \ref{assump:pos},
\ref{assumption:Delta-adjusted-model}/\ref{assumption:control Cox model},
and \ref{asump:consistency} (regularity conditions), we have
\[
\sup_{r}\left\vert \bP\left(n^{1/2}W_{L}^{*}\leq r\mid O_{1:n}\right)-\bP\left\{ n^{1/2}(\hat{\Delta}_{\tau,\mi}-\Delta_{\tau}^{\sen})\leq r\right\} \right\vert \rightarrow0,
\]
in probability, as $n\to\infty$. \end{theorem}

We provide the proof of Theorem \ref{th:WBconsistency} in the Web
Appendix, which draws on the martingale central limit theory \citep{hall1980martingale}
and the asymptotic property of weighted sampling of martingale difference
arrays \citep{pauly2011weighted}. Theorem \ref{th:WBconsistency}
indicates that the distribution of the wild bootstrap statistic consistently
estimates the distribution of the MI estimator.

\section{An application\label{sec:Applications}}
%\begin{itemize}
%\item \textit{Maybe I missed it, but there is no illustration of control-based
%Cox model. }\textbf{\textit{More references are needed to justify
%the comments about control-based imputatio}}\textit{n. Is its appeal
%justified in the application?}
%\item \textit{The application should }\textbf{\textit{allow treatment-specific
%sensitivity analysis parameters }}\textit{unless there is some justification
%for why they should be the same.}
%\end{itemize}
We apply the proposed semiparametric $\delta$-adjusted and control-based Cox model
to a HIV clinical trial. The randomized double-blinded ACTG175 trial
was conducted to compare the treatment effect of a single nucleoside
and two nucleosides in adults with HIV \citep{hammer1996trial}. The
dataset is available in the R package $\mathtt{speff2trial}$. The
event of interest was the progression of the disease defined as the
first occurrence of more than $50$ percent decline in the CD4 cell
count or death. For illustration purposes, we compare the treatment
effect between Zidovudine monotherapy and Zidovudine plus Didanosine
combination therapy in a subgroup of participants who never took any
type of antiretroviral therapy before randomization. In this subgroup,
there were $197$ subjects in the monotherapy group and $185$ subjects
in the combination therapy group. \textcolor{black}{
There are 152 (82.2\%) subjects in Zidovudine plus Didanosine
combination therapy group and 144 (73.0\%) subjects in Zidovudine monotherapy
group censored.
We focus on estimating the RMST
with the truncation time point $24$ months because the ACTG175 study
required at least $24$ months follow-up for subjects.} 
While re-analyzing the data,
we assume CAR in the primary analysis and assume the event times follow
a Cox model adjusting for age, and symptomatic
indicator terms to estimate. The estimated RMST with $95\%$ confidence
interval is $22.1$ $(21.5,22.8)$ months in the monotherapy group
v.s. $23.0$ $(22.6,23.5)$ in the combination therapy. The estimated
between-group RMST difference with $95\%$ confidence interval is
$0.92$ $(0.15,1.68)$. P-value is $0.019$ that indicates a statistically
significant improvement of the combination therapy compared with the
monotherapy.
\textcolor{black}{We also analyze the data using direct estimator of RMST 
\citep{tian2014simple} without imputation using survRM2 package. The results is close to the 
$\delta$-adjusted method when $\delta = 1$, because both methods assume CAR.} However, the
direct estimator does not require  a Cox model for missing data imputation. 

We conduct the sensitivity analysis based on the $\delta$-adjusted 
\textcolor{black}{ and control-based method} 
to evaluate the impact of plausible departures from the CAR
assumption in the primary analysis. \textcolor{black}{One of the main objectives of the ACTG175 trial was to evaluate the additional benefit with the combination therapy on top of Zidovudine. Therefore, we treat the Zidovudine monotherapy group as the control group and the Zidovudine plus Didanosine combination therapy as the test treatment group.} In the sensitivity analysis, we
consider subjects censored before $24$ months as censored for non-administrative
reasons and subjects censored after $24$ months as censored for administrative
reasons. For the imputation models \textcolor{black}{in both $\delta$-adjusted and control-based methods}, we assume CAR for subjects who
were administratively censored in the combination therapy group or
censored in the monotherapy group. 
\textcolor{black}{In $\delta$-adjusted method,} the $\delta$-adjustment is
applied to the primary Cox model for subjects who were non-administratively
censored in the combination therapy group.
{}
The analysis model is the resulting
patten-mixture model carried out by MI \textcolor{black}{with $m=50$}.
We estimate the standard errors by Rubin's combining rule and the
proposed wild bootstrap method \textcolor{black}{with $B = 100$}.

Table \ref{tab:atcg175} summarizes the results. The estimated within-
and between-group standard errors from the wild bootstrap are smaller
than that from Rubin's combining rule for all evaluated methods. This is
coherent with the findings in the simulation study. From the p-value
of each $\delta$, the estimated tipping point of the sensitivity
analysis is larger than $5$ by using wild bootstrap and between $4$
and $5$ by using Rubin's rule. The results from the proposed wild
bootstrap method demonstrate a stronger evidence for the robustness
of the primary analysis compared with the conservative Rubin's rule.
From the sensitivity results based on the wild bootstrap, to eliminate
the statistical significance of the treatment effect, the hazard of
those subjects who were non-administratively censored should be more
than $5$ times higher than subjects with the observed event times
in the same group. 
\textcolor{black}{The control-based method also provides p-values smaller than $0.05$ by using both Wild Bootstrap and Rubin's rule.}
Therefore, the findings from the primary analysis
is robust to the censoring assumption.

\begin{table}
\caption{Analysis of the ACTG175 Trial Data \label{tab:atcg175}}

\vspace{0.25cm}

\centering{}

\resizebox{\textwidth}{!}{%

\begin{tabular}{lcccccccccc}
\hline 
 & \multicolumn{2}{c}{Zidovudine} &  & \multicolumn{4}{c}{Zidovudine plus } & \multicolumn{3}{c}{}\tabularnewline
 & \multicolumn{2}{c}{$(n=197)$} &  & \multicolumn{4}{c}{Didanosine $(n=185)$} & \multicolumn{3}{c}{Difference }\tabularnewline
\hline 
$Method$  & RMST  & SE  &  &  & RMST  & SE  &  & RMST (95\% CI)  & SE  & P-Value\tabularnewline
\hline 
\multicolumn{11}{c}{Primary and Sensitivity Analysis with Wild Bootstrap}\tabularnewline
$\delta=1$  & 22.10  & 0.28  &  &  & 23.04  & 0.22  &  & 0.92 (0.15, 1.68)  & 0.39  & 0.019\tabularnewline
$\delta=2$  & 22.10  & 0.28  &  &  & 23.00  & 0.23 &  & 0.88 (0.11, 1.64)  & 0.39  & 0.024\tabularnewline
$\delta=3$  & 22.10  & 0.28  &  &  & 22.97  & 0.23 &  & 0.84 (0.18, 1.61)  & 0.39  & 0.031\tabularnewline
$\delta=4$  & 22.10  & 0.28  &  &  & 22.93  & 0.23  &  & 0.81 (0.04, 1.58)  & 0.39  & 0.038\tabularnewline
$\delta=5$  & 22.10  & 0.28  &  &  & 22.90  & 0.23  &  & 0.78 (0.02, 1.55)  & 0.39  & 0.047\tabularnewline
Control-based  & 22.12  & 0.31  &  &  & 23.00  & 0.23  &  & 0.88 (0.12, 1.65)  & 0.39  & 0.023\tabularnewline
\hline 
\multicolumn{11}{c}{Primary and Sensitivity Analysis with Rubin's Combining Rule}\tabularnewline
$\delta=1$  & 22.12  & 0.31  &  &  & 23.04  & 0.24  &  & 0.92 (0.14, 1.69)  & 0.39  & 0.020\tabularnewline
$\delta=2$  & 22.12  & 0.31  &  &  & 23.00  & 0.25  &  & 0.88 ( 0.10, 1.67)  & 0.40  & 0.027\tabularnewline
$\delta=3$  & 22.12  & 0.31  &  &  & 22.97  & 0.25  &  & 0.84 ( 0.06, 1.63)  & 0.40  & 0.034\tabularnewline
$\delta=4$  & 22.12  & 0.31  &  &  & 22.93  & 0.26  &  & 0.81 (0.02, 1.60)  & 0.40  & 0.043\tabularnewline
$\delta=5$  & 22.12  & 0.31  &  &  & 22.90  & 0.26  &  & 0.78 ( -0.01, 1.59)  & 0.40  & 0.054\tabularnewline
Control-based  & 22.12  & 0.31  &  &  & 23.00  & 0.25  &  & 0.87 (0.08, 1.65)  & 0.40  & 0.030\tabularnewline
\hline 
\citep{tian2014simple}  & 22.11  & 0.31  &  &  & 23.05  & 0.24  &  & 0.88 (0.11, 1.66)  & 0.40  & 0.026\tabularnewline
\hline 
\multicolumn{11}{l}{In $\delta$-adjusted sensitivity analysis, The value of $\delta$
applied to subjects who were non-administrative censored in the }\tabularnewline
\multicolumn{11}{l}{Zidovudine plus Didanosine group.}\tabularnewline
\end{tabular}} 
\end{table}

\section{Concluding Remarks\label{sec:Concluding-Remarks}}

In this article, we provide a general framework for survival sensitivity
analysis based on \textit{semiparametric $\delta$-adjusted and control-based
Cox models} to assess the impact of plausible departures from CAR.
The $\delta$-adjusted/control-based models are flexible enough to
accommodate different censoring mechanisms by changing the sensitivity
parameter. MI facilitates the use of simple full-sample estimator;
however, the standard Rubin's combining rule\textcolor{black}{{} may be
conservative or anti-conservative when the analysis method is uncongenial
to the imputation model \citep{meng1994multiple,robins2000inference}.
This is likely to occur in our general sensitivity analysis framework
when the full-sample estimator is not an efficient estimator under
the combined data and imputation models. To overcome this issue, \citet{wang1998large}
and \citet{robins2000inference} proposed consistent variance estimators
for imputation estimators in the missing data literature under a parametric
imputation model, which however is not applicable in our survival
sensitivity analysis.} We reformulate the MI estimator as a martingale
series based on the sequential construction of the MI estimator and
propose the wild bootstrap inference based on resampling the martingale
series with a theoretical guarantee for consistency. \textcolor{black}{Although
the new bootstrap procedure is less straightforward than the nonparametric
bootstrap, the increased complexity for implementation can be made
invisible to the practitioners with the SMIM R package. }

\begin{comment}
In such settings, it is now widely recognized that we need to clearly
set out assumptions for the primary analysis, and then explore the
sensitivity of our inferences to analyses under alternative assumptions.
Both primary and sensitivity assumptions need to be relevant and accessible;
this was the motivation for the original work of Carpenter et al,19
where assumptions about postdeviation behaviour of patients were made
by reference to other groups. A further attraction of this approach
is that the primary analysis model is retained in the sensitivity
analysis---being fitted to the imputed data under the sensitivity
scenarios.

\textbf{suggest J2R: }This approach has the advantage that it avoids
what is often a key difficulty in practice---identifying values for
the sensitivity parameters. This difficulty has been widely acknowledged.
For example, Daniels and Hogan47 quote from Scharfstein et al48 who
comment: \textquotedblleft \dots the biggest challenge in conducting
sensitivity analyses is the choice of one or more sensitivity parameterized
functions whose interpretation can be communicated to patient matter
experts with sufficient clarity\dots .\textquotedblright{} It is therefore
encouraging that reference-based sensitivity analysis via multiple
imputation has increasingly been used (see, for example, Philipsen
et al,49 Jans et al,50 Billings et al,51 and Atri et al52) which motivated
us to set out to systematically extend it to the time-to-event setting.
In Section 3, we present a number of possibilities, many derived
\end{comment}
Recently, \citet{cro2019information} suggested that control-based
imputation permits ``\textit{information anchored}'' analysis in
the sense that the information ratio between the analysis with missing
data and the analysis with ``complete'' data is similar for the
primary analysis and the sensitivity analysis. For the longitudinal
continuous data, they showed that standard errors for the primary
analysis and the sensitivity analysis are approximately the same,
and Rubin's combining rule works for information-anchored analysis
when the proportion of missing data is not large. In survival sensitivity
analysis using control-based imputation models, \citet{atkinson2019reference}
showed by simulation that standard error from Rubin's combining rule
is information-anchored in the sense that it increases as proportion
of censored data increases, although the true variance of the MI estimator
decreases. This appears to be ``paradoxical" that the information of
the estimand increases as the missingness rate increases; however,
the true value of the estimand changes with the missingness rate under
the control-based imputation models. So the statistical information
is not required to be increase with the missingness rate. Our inferential
framework targets consistent estimation of the true variance of the
MI estimator. \textit{}

We present the sensitivity analysis framework in the clinical trial
setting, where due to treatment randomization at baseline, the covariate
distribution is balanced between treatment groups. In an observational
study, treatment comparisons may be difficult to make because of confounding.
\citet{chen2001causal} proposed regression-based analysis and \citet{zhang2012double}
proposed weighting-based analysis for the RMST under CAR in observational
studies. In the future, we will extend the proposed SMIM framework
to assess the robustness of study conclusions against CAR in observational
studies.We will also consider other scenarios with additional missing
covariates, repeated measures in longitudinal data, or recurrent event
data \citep{gao2017control}.

\section*{Supporting Information}

The Web Appendix contains technical assumptions, proofs, and additional
simulation results, and the R code that implements the proposed method
is available. 

\bibliographystyle{dcu}
\bibliography{ci}

\pagebreak{} 
\begin{center}
\textbf{\Large{}{}{}{}{}{}Supporting information for ``SMIM:
a unified framework of survival sensitivity analysis using multiple
imputation and martingale'' by Yang et al.}{\Large{} {}{}{}{}{} }{\Large\par}
\par\end{center}

\begin{center}
\par\end{center}

{\Large{}{}{}{}{}{}\pagenumbering{arabic} %reset page counter to 1
\renewcommand*{\thepage}{S\arabic{page}}}{\Large\par}

{\Large{}{}{}{}{}{}\setcounter{lemma}{0} }{\Large\par}

\global\long\def\thelemma{S\arabic{lemma}}%
{\Large{}{}{}{}{}{} }{\Large\par}

{\Large{}{}{}\setcounter{equation}{0}}{\Large\par}

\global\long\def\theequation{S\arabic{equation}}%
{\Large{}{}{}{}{}{} }{\Large\par}

{\Large{}{}{}\setcounter{section}{0} } 
\global\long\def\thesection{S\arabic{section}}%
{\Large{}{}{}{}{}{} }{\Large\par}

\global\long\def\thesubsection{S\arabic{section}.\arabic{subsection}}%

\global\long\def\theassumption{S\arabic{assumption}}%
{\Large{}{}{}{}{}{} }{\Large\par}

{\Large{}{}{}\setcounter{assumption}{0}}{\Large\par}

\global\long\def\thefigure{S\arabic{figure}}%
{\Large{}{}{}{}{}{} }{\Large\par}

{\Large{}{}{}\setcounter{figure}{0}}{\Large\par}

\global\long\def\thetable{S\arabic{table}}%
{\Large{}{}{}{}{}{} }{\Large\par}

{\Large{}{}{}\setcounter{table}{0}}{\Large\par}

Section~S1 provides the preliminary for the proofs.
Section~S2 establishes the asymptotic linearization of $\hat{S}_{0,\mi}(t)$. 
Section S3 describes the $\sigma-$fields. 
Sections S4 and S5
provide the proofs of Theorem \ref{th:MI-point} and Theorem \ref{th:WBconsistency}.
Section S6  presents a comprehensive simulation study.

\section{Preliminary \label{sec:Preliminary}}

We adopt the counting process theory of \citet{andersen1982cox} in
our theoretical framework. We state the existing results which will
be used in our proof throughout.

To simplify the exposition, we introduce additional notation. We use
$\overset{p}{\to}$ and $\overset{d}{\to}$ to represent ``converge
in probability as $n\to\infty$" and ``converge in distribution
as $n\to\infty$", respectively. Also, let $n_{1}/n\rightarrow p_{1}\in(0,1)$
and $n_{0}/n\rightarrow p_{0}\in(0,1)$, as $n\rightarrow\infty$.
We do not state this condition formally as an assumption because it
holds trivially for most of clinical trials where the two treatment
groups are relatively balanced in their sample sizes.

Let $X_{i}^{\otimes l}$ denote $1$ for $l=0,$ $X_{i}$ for $l=1$,
and $X_{i}X_{i}^{\T}$ for $l=2$. Define 
\[
U_{a}^{(l)}(\beta_{a},t)=\frac{1}{n_{a}}\sum_{i=1}^{n}\bone(A_{i}=a)X_{i}^{\otimes l}e^{\beta_{a}^{\T}X_{i}}Y_{i}(t),\ u_{a}^{(l)}(\beta_{a},t)=\E\left\{ X^{\otimes l}e^{\beta_{a}^{\T}X}Y(t)\right\} ,
\]
where $u_{a}^{(l)}(\beta_{a},t)$ is the expectation of $U_{a}^{(l)}(\beta_{a},t)$,
for $l=0,1,2$. Moreover, define

\[
E_{a}(\beta_{a},t)=\frac{U_{a}^{(1)}(\beta_{a},t)}{U_{a}^{(0)}(\beta_{a},t)},\ {\color{black}e_{a}(\beta_{a},t)=\frac{u_{a}^{(1)}(\beta_{a},t)}{u_{a}^{(0)}(\beta_{a},t)}.}
\]
The maximum partial likelihood estimator $\hat{\beta}_{a}$ solves
\[
\mathcal{S}_{a,n}(\beta_{a})=\frac{1}{n_{a}}\sum_{i=1}^{n}\bone(A_{i}=a)\int_{0}^{\tau}\left\{ X_{i}-\frac{U_{1}^{(1)}(\beta_{a},u)}{U_{1}^{(0)}(\beta_{a},u)}\right\} \de N_{i}(u)=0.
\]

We state the standard asymptotic results for $\hat{\beta}_{a}$ and
$\hat{\lambda}_{a}(\cdot)$ requiring certain regularity conditions.
To avoid too many technical distractions, we omit the exact conditions
in Assumption \ref{asump:consistency} for the consistency and uniform
convergency of the estimators of Cox models.

\begin{assumption}{\label{asump:consistency}} 
	i) (Positivity)	There exists a constant $c$ such that with probability one, $S_{a}(t\mid X_{i})\geq c>0$
	for $t$ in $[0,\tau]$ and $a=0,1$.
	ii) Conditions A--D
in \citet{andersen1982cox} hold for treatment group $a=0,1.$

\end{assumption}

Following \citet{andersen1982cox}, we have 
\begin{eqnarray}
n_{a}^{1/2}(\hat{\beta}_{a}-\beta_{a}) & = & \Gamma_{a}^{-1}\frac{1}{n_{a}^{1/2}}\sum_{i=1}^{n}\bone(A_{i}=a)H_{a,i}+o_{p}(1),\label{eq:beta}
\end{eqnarray}
where $\Gamma_{a}=\E\{-\partial\mathcal{S}_{a,n}(\beta_{a})/\partial\beta_{a}^{\T}\}$
is the Fisher information matrix of $\beta_{a}$, $H_{a,i}=\int_{0}^{L}\{X_{i}-e_{a}(\beta_{a},u)\}\bone(A_{i}=a)\de M_{a,i}(u)$,
and 
\begin{equation}
\de M_{a,i}(t)=\de N_{i}(u)-e^{\beta_{a}^{\T}X_{i}}Y_{i}(u)\lambda_{a,0}(u)\de u.\label{eq:dM}
\end{equation}
Moreover, $n^{1/2}\{S_{a}(t\mid X_{i};\hat{\theta})-S_{a}(t\mid X_{i})\}$
converges uniformly to a Gaussian process in $[0,L]$ for all $X_{i}$.

\section{Asymptotic linearization of $\hat{S}_{0,\mi}(t)$}

To obtain the asymptotic linearization of $\hat{S}_{0,\mi}(t)$, we
have 
\begin{align}
 & n^{1/2}\left\{ \hat{S}_{0,\mi}(t)-S_{0}^{\sen}(t)\right\} \nonumber \\
 & =\frac{n^{1/2}}{mn_{0}}\sum_{j=1}^{m}\sum_{i=1}^{n}(1-A_{i})\{1-Y_{i}(t)\}\left[\bone(T_{i}^{*(j)}\geq t)-S_{0}\{t\mid H_{i}(t);\hat{\theta}\}\right]\label{eq:MI0-1}\\
 & +\frac{n^{1/2}}{n_{0}}\sum_{i=1}^{n}(1-A_{i})\left[Y_{i}(t)+\{1-Y_{i}(t)\}(1-I_{i})S_{0}\{t\mid H_{i}(t);\theta\}-S_{0}^{\sen}(t)\right]\label{eq:MI0-2}\\
 & +\frac{n^{1/2}}{n_{0}}\sum_{i=1}^{n}(1-A_{i})\phi_{0,i}(t)+o_{p}(1),\label{eq:MI0-3}
\end{align}
where the exact expression of $\phi_{0,i}(t)$ is given in Section
\ref{sec:Proof-of-Thm1}, reflecting the estimation of $\{\lambda_{0}(\cdot),\beta_{0}\}$.
In our context, the imputation for the control group uses the information
only from the control group. By the imputation and estimation procedures,
(\ref{eq:MI0-1})--(\ref{eq:MI0-3}) have (conditional) mean zero.

\section{$\sigma$-fields for the martingales}

We consider the $\sigma$-fields as follows 
\[
\F_{n,k}=\begin{cases}
\sigma\left(O_{1},\ldots,O_{k}\right), & \text{for }k=i\ \ \ \ (1\leq i\leq n_{1}),\\
\sigma\left(O_{1},\ldots,O_{n_{1}},T_{1}^{*(1)},\ldots,T_{i}^{*(j)}\right), & \text{for }k=n_{1}+(i-1)m+j\\
 & \ \ \ \ (1\leq i\leq n_{1},1\leq j\leq m),\\
\sigma\left(O_{1},\ldots,O_{n_{1}},T_{1}^{*(1)},\ldots,T_{n_{1}}^{*(m)}\right., & \text{for }k=(1+m)n_{1}+i\\
\ \ \ \ \left.O_{n_{1}+1},\ldots,O_{k}\vphantom{T_{n_{1}}^{*(m)}}\right), & \ \ \ \ (n_{1}+1\leq i\leq n),\\
\sigma\left(O_{1},\ldots,O_{n_{1}},T_{1}^{*(1)},\ldots,T_{n_{1}}^{*(m)},\right. & \text{for }k=(1+m)n_{1}+n_{0}+(i-1)m+j\\
\ \ \ \ \left.O_{n_{1}+1},\ldots,O_{n},T_{n_{1}+1}^{*(1)},\ldots,T_{i}^{*(j)}\right), & \ \ \ \ (n_{1}+1\leq i\leq n,1\leq j\leq m).
\end{cases}
\]

\section{Proof of Theorem \ref{th:MI-point} \label{sec:Proof-of-Thm1}}

We first derive the martingale representation of the MI estimator
under $\delta$-adjusted Cox models and control-based Cox models,
separately. Then, we apply the martingale CLT to derive the asymptotic
distribution of the MI estimator.

\subsection{Delta-adjusted Cox models}

A key step is to separate the imputation step and the estimation step.
We start with treatment group $a=1$. For the imputations, it is important
to recognize that $T_{i}^{*(j)}$ follows a time-dependent Cox model
with the conditional survival function $S_{1}\{t\mid H_{i}(t);\hat{\theta}\}$
for $t>U_{i},$ where 
\begin{eqnarray*}
S_{1}\{t\mid H_{i}(t);\theta\} & = & \begin{cases}
\exp\left\{ -\int_{U_{i}}^{t}\lambda_{1}(u)e^{\beta_{1}^{\T}X_{i}}\de u\right\} , & \text{if }A_{i}=1,R_{i}=1,\\
\exp\left\{ -\delta\int_{U_{i}}^{t}\lambda_{0}(u)e^{\beta_{0}^{\T}X_{i}}\de u\right\} , & \text{if }A_{i}=1,R_{i}=2.
\end{cases}
\end{eqnarray*}

We express the MI estimator of $S_{1}^{\delta\text{-adj}}(t)$ as
\begin{eqnarray}
 &  & n^{1/2}\left\{ \hat{S}_{1,\mi}(t)-S_{1}^{\delta\text{-adj}}(t)\right\} \nonumber \\
 & = & \frac{n^{1/2}}{mn_{1}}\sum_{j=1}^{m}\sum_{i=1}^{n}A_{i}\{\bone(T_{i}^{*(j)}\geq t)-S_{1}^{\delta\text{-adj}}(t)\}\nonumber \\
 & = & \frac{n^{1/2}}{mn_{1}}\sum_{j=1}^{m}\sum_{i=1}^{n}A_{i}[\bone(T_{i}^{*(j)}\geq t)-S_{1}\{t\mid H_{i}(t);\hat{\theta}\}]+\frac{n^{1/2}}{n_{1}}\sum_{i=1}^{n}A_{i}[S_{1}\{t\mid H_{i}(t);\hat{\theta}\}-S_{1}^{\delta\text{-adj}}(t)]\nonumber \\
 & = & \frac{n^{1/2}}{mn_{1}}\sum_{j=1}^{m}\sum_{i=1}^{n}A_{i}\{1-Y_{i}(t)\}[\bone(T_{i}^{*(j)}\geq t)-S_{1}\{t\mid H_{i}(t);\hat{\theta}\}]\label{eq:term1}\\
 &  & +\frac{n^{1/2}}{n_{1}}\sum_{i=1}^{n}\left[A_{i}Y_{i}(t)+A_{i}\{1-Y_{i}(t)\}(1-I_{i})S_{1}\{t\mid H_{i}(t);\hat{\theta}\}-S_{1}^{\delta\text{-adj}}(t)\right]+o_{p}(1),\label{eq:term3}
\end{eqnarray}
where (\ref{eq:term1}) follows because $\bone(T_{i}^{*(j)}\geq t)-S_{1}\{t\mid H_{i}(t);\hat{\theta}\}=0$
for subject $i$ with $\{A_{i}=1,$$Y_{i}(t)=1\}$, and (\ref{eq:term3})
follows because $A_{i}\{1-Y_{i}(t)\}I_{i}S_{1}\{t\mid H_{i}(t);\hat{\theta}\}=0$.

By the counting process theory, we can express the term $n_{1}^{-1/2}\sum_{i=1}^{n}A_{i}\{1-Y_{i}(t)\}(1-I_{i})S_{1}\{t\mid H_{i}(t);\hat{\theta}\}$
in (\ref{eq:term3}) further as {\footnotesize{}
\begin{eqnarray}
 &  & \frac{1}{n_{1}^{1/2}}\sum_{i=1}^{n}A_{i}\{1-Y_{i}(t)\}(1-I_{i})S_{1}\{t\mid H_{i}(t);\hat{\theta}\}\nonumber \\
 & = & \frac{1}{n_{1}^{1/2}}\sum_{i=1}^{n}A_{i}\{1-Y_{i}(t)\}(1-I_{i})\exp\left\{ -\int_{U_{i}}^{t}\hat{\lambda}_{1}(u)\delta^{\bone(R_{i}=2)}e^{\hat{\beta}_{1}^{\T}X_{i}}\de u\right\} \nonumber \\
 & = & \frac{1}{n_{1}^{1/2}}\sum_{i=1}^{n}A_{i}\{1-Y_{i}(t)\}(1-I_{i})S_{1}\{t\mid H_{i}(t);\theta\}\nonumber \\
 &  & +\frac{1}{n_{1}^{1/2}}\sum_{i=1}^{n}A_{i}\{1-Y_{i}(t)\}(1-I_{i})S_{1}\{t\mid H_{i}(t);\theta\}\left[-\int_{U_{i}}^{t}\delta^{\bone(R_{i}=2)}e^{\beta_{1}^{\T}X_{i}}\left\{ \hat{\lambda}_{1}(u)-\lambda_{1}(u)\right\} \de u\right]\label{eq:term4}\\
 &  & +\left[\frac{1}{n_{1}}\sum_{i=1}^{n}A_{i}\{1-Y_{i}(t)\}(1-I_{i})S_{1}\{t\mid H_{i}(t);\theta\}\left\{ -\int_{U_{i}}^{t}\lambda_{1}(u)\delta^{\bone(R_{i}=2)}e^{\beta_{1}^{\T}X_{i}}X_{i}\de u\right\} \right]\label{eq:term6}\\
 &  & \times n_{1}^{1/2}\left(\hat{\beta}_{1}-\beta_{1}\right).
\end{eqnarray}
}For (\ref{eq:term4}), we further express the key term as 
\begin{eqnarray}
 &  & \int_{U_{i}}^{t}\delta^{\bone(R_{i}=2)}e^{\beta_{1}^{\T}X_{i}}\left\{ \hat{\lambda}_{1}(u)-\lambda_{1}(u)\right\} \de u\nonumber \\
 & = & \int_{U_{i}}^{t}\delta^{\bone(R_{i}=2)}e^{\beta_{1}^{\T}X_{i}}\left\{ \frac{n_{1}^{-1}\sum_{j=1}^{n}A_{j}\de N_{j}(u)}{U_{1}^{(0)}(\hat{\beta}_{1},u)}-\frac{n_{1}^{-1}\sum_{j=1}^{n}A_{j}\de N_{j}(u)}{U_{1}^{(0)}(\beta_{1},u)}\right\} \nonumber \\
 &  & +\int_{U_{i}}^{t}\delta^{\bone(R_{i}=2)}e^{\beta_{1}^{\T}X_{i}}\left\{ \frac{n_{1}^{-1}\sum_{j=1}^{n}A_{j}\de N_{j}(u)}{U_{1}^{(0)}(\beta_{1},u)}-\lambda_{1}(u)\de u\right\} \nonumber \\
 & = & -\left[\int_{C_{i}}^{t}\delta^{\bone(R_{i}=2)}e^{\beta_{1}^{\T}X_{i}}\frac{U_{1}^{(1)}(\beta_{1},u)}{\left\{ U_{1}^{(0)}(\beta_{1},u)\right\} ^{2}}\left\{ n_{1}^{-1}\sum_{j=1}^{n}\de N_{j}(u)\right\} \right]^{\T}\left(\hat{\beta}_{1}-\beta_{1}\right)\nonumber \\
 &  & +\int_{U_{i}}^{t}\delta^{\bone(R_{i}=2)}e^{\beta_{1}^{\T}X_{i}}\frac{n_{1}^{-1}\sum_{j=1}^{n}A_{j}\de M_{1,j}(u)}{U_{1}^{(0)}(\beta_{1},u)}+o_{p}(1)\nonumber \\
 & = & -\left\{ \int_{U_{i}}^{t}\delta^{\bone(R_{i}=2)}e^{\beta_{1}^{\T}X_{i}}e_{1}(\beta_{1},u)\lambda_{1}(u)\de u\right\} ^{\T}\left(\hat{\beta}_{1}-\beta_{1}\right)\nonumber \\
 &  & +\int_{U_{i}}^{t}\delta^{\bone(R_{i}=2)}e^{\beta_{1}^{\T}X_{i}}\frac{n_{1}^{-1}\sum_{j=1}^{n}A_{j}\de M_{1,j}(u)}{U_{1}^{(0)}(\beta_{1},u)}+o_{p}(1),\label{eq:term5}
\end{eqnarray}
where $\de M_{1,j}(u)$ is defined in (\ref{eq:dM}). Denote {\small{}
\begin{eqnarray*}
g_{a,0}(t) & = & \E\left[\bone(A_{i}=a)\{1-Y_{i}(t)\}(1-I_{i})S_{a}\{t\mid H_{i}(t);\theta\}\delta^{\bone(R_{i}=2)}e^{\beta_{a}^{\T}X_{i}}\right],\\
g_{a,1}(t) & = & \E\left[\bone(A_{i}=a)\{1-Y_{i}(t)\}(1-I_{i})S_{a}\{t\mid H_{i}(t);\theta\}\left\{ \int_{U_{i}}^{t}\delta^{\bone(R_{i}=2)}e^{\beta_{a}^{\T}X_{i}}X_{i}\lambda_{a}(u)\de u\right\} \right],\\
g_{a,2}(t) & = & \E\left[\bone(A_{i}=a)\{1-Y_{i}(t)\}(1-I_{i})S_{a}\{t\mid H_{i}(t);\theta\}\left\{ \int_{U_{i}}^{t}\delta^{\bone(R_{i}=2)}e^{\beta_{a}^{\T}X_{i}}e_{a}(\beta_{a},u)\lambda_{a}(u)\de u\right\} \right],
\end{eqnarray*}
}for $a=0,1$.

Plugging (\ref{eq:term5}) in (\ref{eq:term4}) becomes 
\begin{eqnarray}
 &  & \frac{1}{n_{1}^{1/2}}\sum_{i=1}^{n_{1}}A_{i}\{1-Y_{i}(t)\}(1-I_{i})S_{1}\{t\mid H_{i}(t);\hat{\theta}\}\nonumber \\
 & = & \frac{1}{n_{1}^{1/2}}\sum_{i=1}^{n_{1}}A_{i}\{1-Y_{i}(t)\}(1-I_{i})S_{1}\{t\mid H_{i}(t);\theta\}\nonumber \\
 &  & +\left\{ g_{1,2}(t)-g_{1,1}(t)\right\} ^{\T}n_{1}^{1/2}\left(\hat{\beta}_{1}-\beta_{1}\right)-n_{1}^{-1/2}\sum_{j=1}^{n}\int_{U_{j}}^{t}\frac{g_{1,0}(u)}{s_{0}(\beta_{1},u)}A_{j}\de M_{1,j}(u)+o_{p}(1)\nonumber \\
 & = & \frac{1}{n_{1}^{1/2}}\sum_{i=1}^{n_{1}}A_{i}\{1-Y_{i}(t)\}(1-I_{i})S_{1}\{t\mid H_{i}(t);\theta\}\nonumber \\
 &  & +\frac{1}{n_{1}^{1/2}}\sum_{i=1}^{n_{1}}\left[\left\{ g_{1,2}(t)-g_{1,1}(t)\right\} ^{\T}\Gamma_{1}^{-1}A_{i}H_{1,i}-\int_{U_{i}}^{t}\frac{g_{1,0}(u)}{s_{0}(\beta_{1},u)}A_{i}\de M_{1,i}(u)\right]+o_{p}(1),\label{eq:term2}
\end{eqnarray}
where the second equality follows by (\ref{eq:beta}).

Combining (\ref{eq:term1}) and (\ref{eq:term2}) leads to 
\begin{eqnarray}
 &  & n^{1/2}\left\{ \hat{S}_{1,\mi}(t)-S_{1}^{\delta\text{-adj}}(t)\right\} \nonumber \\
 & = & \frac{n^{1/2}}{mn_{1}}\sum_{j=1}^{m}\sum_{i=1}^{n_{1}}\left[A_{i}\{1-Y_{i}(t)\}\{\bone(T_{i}^{*(j)}\geq t)-S_{1}(t\mid O_{i};\hat{\theta}_{1})\}\right]\nonumber \\
 &  & +\frac{n^{1/2}}{n_{1}}\sum_{i=1}^{n_{1}}A_{i}\left[\phi_{11,i}(t)+Y_{i}(t)+\{1-Y_{i}(t)\}(1-I_{i})S_{1}\{t\mid H_{i}(t);\theta\}-S_{1}^{\delta\text{-adj}}(t)\right]\label{eq:S1}\\
 &  & +o_{p}(1).\nonumber 
\end{eqnarray}
where 
\begin{eqnarray}
\phi_{11,i}(t) & = & \left\{ g_{1,2}(t)-g_{1,1}(t)\right\} ^{\T}\Gamma_{1}^{-1}H_{1,i}-\int_{U_{i}}^{t}\frac{g_{1,0}(u)}{u_{0}(\beta_{1},u)}\de M_{1,i}(u).\label{eq:phi11}
\end{eqnarray}

Similarly, for treatment group $a=0$, define 
\begin{equation}
\phi_{0,i}(t)=\left\{ g_{0,2}(t)-g_{0,1}(t)\right\} ^{\T}\Gamma_{0}^{-1}H_{0,i}-\int_{U_{i}}^{t}\frac{g_{0,0}(u)}{u_{0}(\beta_{0},u)}\de M_{0,i}(u),\label{eq:phi0}
\end{equation}
We have {\small{}
\begin{eqnarray}
 &  & n^{1/2}\left\{ \hat{S}_{0,\mi}(t)-S_{0}^{\delta\text{-adj}}(t)\right\} \nonumber \\
 & = & \frac{n^{1/2}}{mn_{0}}\sum_{j=1}^{m}\sum_{i=1}^{n}(1-A_{i})\{1-Y_{i}(t)\}[\bone(T_{i}^{*(j)}\geq t)-S_{0}\{t\mid H_{i}(t);\hat{\theta}\}]\nonumber \\
 &  & +\frac{n^{1/2}}{n_{0}}\sum_{i=1}^{n}(1-A_{i})\left\{ \phi_{0,i}(t)+Y_{i}(t)+\{1-Y_{i}(t)\}(1-I_{i})S_{0}\{t\mid H_{i}(t);\theta\}-S_{0}^{\delta\text{-adj}}(t)\right\} \label{eq:S0}\\
 &  & +o_{p}(1).\nonumber 
\end{eqnarray}
}{\small\par}

The martingale series approximation of $\hat{\Delta}_{\tau,\mi}$
follows by plugging (\ref{eq:S1}) and (\ref{eq:S0}) into 
\begin{eqnarray*}
n^{1/2}\left(\hat{\Delta}_{\tau,\mi}-\Delta_{\tau}^{\delta\text{-adj}}\right) & = & n^{1/2}\left[\Psi_{\tau}\{\hat{S}_{1,\mi}(t),\hat{S}_{0,\mi}(t)\}-\Delta_{\tau}^{\delta\text{-adj}}\right]\\
 & = & \sum_{a=0}^{1}\int_{0}^{\tau}\psi_{a}(t)\left\{ \hat{S}_{a,\mi}(t)-S_{a}(t)\right\} \de t+o_{p}(1)=\sum_{k=1}^{(1+m)n}\xi_{n,k}+o_{p}(1),
\end{eqnarray*}
where the $\xi_{n,k}$ terms are given in (\ref{eq:xi1}) with $\phi_{10,i}(t)=0$
and $\phi_{11,i}(t)$ and $\phi_{0,i}(t)$ given in (\ref{eq:phi11})
and (\ref{eq:phi0}), respectively.

\subsection{Control-based Cox models}

We focus on the treatment group $a=1$. Under the control-based imputation
model, the MI estimator $\hat{S}_{1,\mi}(t)$ depends on not only
the parameter estimator in the treatment group but also the parameter
estimator in the control group. Following the same steps for (\ref{eq:term3}),
we express the MI estimator as 
\begin{eqnarray}
 &  & n^{1/2}\left\{ \hat{S}_{1,\mi}(t)-S_{1}^{\delta\text{-cb}}(t)\right\} \nonumber \\
 & = & \frac{n^{1/2}}{mn_{1}}\sum_{j=1}^{m}\sum_{i=1}^{n}A_{i}\{1-Y_{i}(t)\}[\bone(T_{i}^{*(j)}\geq t)-S_{1}\{t\mid H_{i}(t);\hat{\theta}\}]\label{eq:1}\\
 &  & +\frac{n^{1/2}}{n_{1}}\sum_{i=1}^{n}A_{i}\left[Y_{i}(t)+\{1-Y_{i}(t)\}(1-I_{i})S_{1}\{t\mid H_{i}(t);\hat{\theta}\}-S_{1}^{\delta\text{-cb}}(t)\right],\label{eq:2}
\end{eqnarray}
where under the imputation based on the control-based Cox model, 
\begin{eqnarray*}
S_{1}\{t\mid H_{i}(t);\theta\} & = & \begin{cases}
\exp\left\{ -\int_{U_{i}}^{t}\lambda_{1}(u)e^{\beta_{1}^{\T}X_{i}}\de u\right\} , & \text{if }A_{i}=1,R_{i}=1,\\
\exp\left\{ -\delta\int_{U_{i}}^{t}\lambda_{0}(u)e^{\beta_{0}^{\T}X_{i}}\de u\right\} , & \text{if }A_{i}=1,R_{i}=2,
\end{cases}
\end{eqnarray*}
for $t\geq U_{i}$.

By the counting process theory, we can further express $n^{1/2}n_{1}^{-1}\sum_{i=1}^{n}A_{i}\{1-Y_{i}(t)\}(1-I_{i})S_{1}\{t\mid H_{i}(t);\hat{\theta}\}$
in (\ref{eq:2}) as 
\begin{eqnarray}
 &  & \frac{n^{1/2}}{n_{1}}\sum_{i=1}^{n}A_{i}\{1-Y_{i}(t)\}(1-I_{i})S_{1}\{t\mid H_{i}(t);\hat{\theta}\}\nonumber \\
 & = & \frac{n^{1/2}}{n_{1}}\sum_{i=1}^{n}A_{i}\{1-Y_{i}(t)\}(1-I_{i})\bone(R_{i}=1)\exp\left\{ -\int_{U_{i}}^{t}\hat{\lambda}_{1}(u)e^{\hat{\beta}_{1}^{\T}X_{i}}\de u\right\} \nonumber \\
 &  & +\frac{n^{1/2}}{n_{1}}\sum_{i=1}^{n}A_{i}\{1-Y_{i}(t)\}(1-I_{i})\bone(R_{i}=2)\exp\left\{ -\delta\int_{U_{i}}^{t}\hat{\lambda}_{0}(u)e^{\hat{\beta}_{0}^{\T}X_{i}}\de u\right\} \nonumber \\
 & = & \frac{n^{1/2}}{n_{1}}\sum_{i=1}^{n}A_{i}\{1-Y_{i}(t)\}(1-I_{i})\bone(R_{i}=1)\exp\left\{ -\int_{U_{i}}^{t}\lambda_{1}(u)e^{\beta_{1}^{\T}X_{i}}\de u\right\} \nonumber \\
 &  & +\frac{n^{1/2}}{n_{1}}\sum_{i=1}^{n}A_{i}\{1-Y_{i}(t)\}(1-I_{i})\bone(R_{i}=2)\exp\left\{ -\delta\int_{U_{i}}^{t}\lambda_{0}(u)e^{\beta_{0}^{\T}X_{i}}\de u\right\} \nonumber \\
 &  & +\frac{n^{1/2}}{n_{1}}\sum_{i=1}^{n}A_{i}\{1-Y_{i}(t)\}(1-I_{i})\bone(R_{i}=1)S_{1}\{t\mid H_{i}(t);\theta\}\left[-\int_{U_{i}}^{t}e^{\beta_{1}^{\T}X_{i}}\left\{ \hat{\lambda}_{1}(u)-\lambda_{1}(u)\right\} \de u\right]\nonumber \\
 &  & +\left[\frac{1}{n_{1}}\sum_{i=1}^{n}A_{i}\{1-Y_{i}(t)\}(1-I_{i})\bone(R_{i}=1)S_{1}\{t\mid H_{i}(t);\theta\}\left\{ -\int_{U_{i}}^{t}\lambda_{1}(u)e^{\beta_{1}^{\T}X_{i}}X_{i}\de u\right\} \right]\nonumber \\
 &  & \times n^{1/2}\left(\hat{\beta}_{1}-\beta_{1}\right)\nonumber \\
 &  & +\frac{n^{1/2}}{n_{1}}\sum_{i=1}^{n}A_{i}\{1-Y_{i}(t)\}(1-I_{i})\bone(R_{i}=2)S_{1}\{t\mid H_{i}(t);\theta\}\left[-\delta\int_{U_{i}}^{t}e^{\beta_{0}^{\T}X_{i}}\left\{ \hat{\lambda}_{0}(u)-\lambda_{0}(u)\right\} \de u\right]\nonumber \\
 &  & +\left[\frac{1}{n_{1}}\sum_{i=1}^{n}A_{i}\{1-Y_{i}(t)\}(1-I_{i})\bone(R_{i}=2)S_{1}\{t\mid H_{i}(t);\theta\}\left\{ -\delta\int_{U_{i}}^{t}\lambda_{0}(u)e^{\beta_{0}^{\T}X_{i}}X_{i}\de u\right\} \right]\nonumber \\
 &  & \times n^{1/2}\left(\hat{\beta}_{0}-\beta_{0}\right).\label{eq:3}
\end{eqnarray}
Denote 
\begin{eqnarray*}
\tilde{g}_{1,0}(t) & = & \E\left[A_{i}\{1-Y_{i}(t)\}(1-I_{i})\bone(R_{i}=1)S_{1}\{t\mid H_{i}(t);\theta\}e^{\beta_{1}^{\T}X_{i}}\right],\\
\tilde{g}_{1,1}(t) & = & \E\left[A_{i}\{1-Y_{i}(t)\}(1-I_{i})\bone(R_{i}=1)S_{1}\{t\mid H_{i}(t);\theta\}\left\{ \int_{U_{i}}^{t}\lambda_{1}(u)e^{\beta_{1}^{\T}X_{i}}X_{i}^{\T}\de u\right\} \right],\\
\tilde{g}_{1,2}(t) & = & \E\left[A_{i}\{1-Y_{i}(t)\}(1-I_{i})\bone(R_{i}=1)S_{1}\{t\mid H_{i}(t);\theta\}\left\{ \int_{U_{i}}^{t}\lambda_{1}(u)e^{\beta_{1}^{\T}X_{i}}e_{1}(\beta_{1},u)^{\T}\de u\right\} \right],\\
\tilde{g}_{0,0}(t) & = & \E\left[A_{i}\{1-Y_{i}(t)\}(1-I_{i})\bone(R_{i}=2)S_{1}\{t\mid H_{i}(t);\theta\}\delta e^{\beta_{0}^{\T}X_{i}}\right],\\
\tilde{g}_{0,1}(t) & = & \E\left[A_{i}\{1-Y_{i}(t)\}(1-I_{i})\bone(R_{i}=2)S_{1}\{t\mid H_{i}(t);\theta\}\left\{ \delta\int_{U_{i}}^{t}\lambda_{0}(u)e^{\delta\beta_{0}^{\T}X_{i}}X_{i}^{\T}\de u\right\} \right],\\
\tilde{g}_{0,2}(t) & = & \E\left[A_{i}\{1-Y_{i}(t)\}(1-I_{i})\bone(R_{i}=2)S_{1}\{t\mid H_{i}(t);\theta\}\left\{ \delta\int_{U_{i}}^{t}\lambda_{0}(u)e^{\beta_{0}^{\T}X_{i}}e_{0}(\beta_{0},u)^{\T}\de u\right\} \right].
\end{eqnarray*}
Then, we can express (\ref{eq:3}) further as 
\begin{eqnarray}
 & = & \frac{n^{1/2}}{n_{1}}\sum_{i=1}^{n_{1}}A_{i}\{1-Y_{i}(t)\}(1-I_{i})\bone(R_{i}=1)\exp\left\{ -\int_{U_{i}}^{t}\lambda_{1}(u)e^{\beta_{1}^{\T}X_{i}}\de u\right\} \nonumber \\
 &  & +\frac{n^{1/2}}{n_{1}}\sum_{i=1}^{n_{1}}A_{i}\{1-Y_{i}(t)\}(1-I_{i})\bone(R_{i}=2)\exp\left\{ -\delta\int_{U_{i}}^{t}\lambda_{0}(u)e^{\beta_{0}^{\T}X_{i}}\de u\right\} \nonumber \\
 &  & +\left\{ \tilde{g}_{1,2}(t)-\tilde{g}_{1,1}(t)\right\} ^{\T}n^{1/2}\left(\hat{\beta}_{1}-\beta_{1}\right)-\frac{n^{1/2}}{n_{1}}\sum_{j=1}^{n}\int_{U_{i}}^{t}\frac{\tilde{g}_{1,0}(u)}{s_{0}(\beta_{1},u)}A_{j}\de M_{1,j}(u)\nonumber \\
 &  & +\left\{ \tilde{g}_{0,2}(t)-\tilde{g}_{0,1}(t)\right\} ^{\T}n^{1/2}\left(\hat{\beta}_{0}-\beta_{0}\right)-\frac{n^{1/2}}{n_{1}}\sum_{j=1}^{n}\int_{U_{i}}^{t}\frac{\tilde{g}_{0,0}(u)}{s_{0}(\beta_{0},u)}(1-A_{j})\de M_{0,j}(u)+o_{p}(1)\nonumber \\
 & = & \frac{n^{1/2}}{n_{1}}\sum_{i=1}^{n_{1}}A_{i}\{1-Y_{i}(t)\}(1-I_{i})\bone(R_{i}=1)\exp\left\{ -\int_{U_{i}}^{t}\lambda_{1}(u)e^{\beta_{1}^{\T}X_{i}}\de u\right\} \nonumber \\
 &  & +\frac{n^{1/2}}{n_{1}}\sum_{i=1}^{n_{1}}A_{i}\{1-Y_{i}(t)\}(1-I_{i})\bone(R_{i}=2)\exp\left\{ -\delta\int_{U_{i}}^{t}\lambda_{0}(u)e^{\beta_{0}^{\T}X_{i}}\de u\right\} \nonumber \\
 &  & +\frac{n^{1/2}}{n_{1}}\sum_{i=1}^{n}\left[\left\{ \tilde{g}_{1,2}(t)-\tilde{g}_{1,1}(t)\right\} ^{\T}\Gamma_{1}^{-1}A_{i}H_{1,i}-\int_{U_{i}}^{t}\frac{\tilde{g}_{1,0}(u)}{s_{0}(\beta_{1},u)}A_{i}\de M_{1,i}(u)\right]\nonumber \\
 &  & +\frac{n^{1/2}}{n_{0}}\sum_{i=1}^{n}\left[\left\{ \tilde{g}_{0,2}(t)-\tilde{g}_{0,1}(t)\right\} ^{\T}\Gamma_{0}^{-1}(1-A_{i})H_{0,i}-\int_{U_{i}}^{t}\frac{\tilde{g}_{0,0}(u)}{s_{0}(\beta_{0},u)}(1-A_{i})\de M_{0,i}(u)\right]+o_{p}(1)\nonumber \\
 & = & \frac{n^{1/2}}{n_{1}}\sum_{i=1}^{n_{1}}A_{i}\{1-Y_{i}(t)\}(1-I_{i})\bone(R_{i}=1)\exp\left\{ -\int_{U_{i}}^{t}\lambda_{1}(u)e^{\beta_{1}^{\T}X_{i}}\de u\right\} \nonumber \\
 &  & +\frac{n^{1/2}}{n_{1}}\sum_{i=1}^{n_{1}}A_{i}\{1-Y_{i}(t)\}(1-I_{i})\bone(R_{i}=2)\exp\left\{ -\delta\int_{U_{i}}^{t}\lambda_{0}(u)e^{\beta_{0}^{\T}X_{i}}\de u\right\} \nonumber \\
 &  & +\frac{n^{1/2}}{n_{1}}\sum_{i=1}^{n}A_{i}\left[\left\{ \tilde{g}_{1,2}(t)-\tilde{g}_{1,1}(t)\right\} ^{\T}\Gamma_{1}^{-1}H_{1,i}-\int_{U_{i}}^{t}\frac{\tilde{g}_{1,0}(u)}{s_{0}(\beta_{1},u)}\de M_{1,i}(u)\right]\nonumber \\
 &  & +\frac{n^{1/2}}{n_{0}}\sum_{i=1}^{n}(1-A_{i})\left[\left\{ \tilde{g}_{0,2}(t)-\tilde{g}_{0,1}(t)\right\} ^{\T}\Gamma_{0}^{-1}H_{0,i}-\int_{U_{i}}^{t}\frac{\tilde{g}_{0,0}(u)}{s_{0}(\beta_{0},u)}\de M_{0,i}(u)\right]+o_{p}(1).\label{eq:4}
\end{eqnarray}
Combining (\ref{eq:1}) and (\ref{eq:4}) leads to 
\begin{eqnarray*}
 &  & n^{1/2}\left\{ \hat{S}_{1,\mi}(t)-S_{1}^{\delta\text{-cb}}(t)\right\} \\
 & = & \frac{n^{1/2}}{mn_{1}}\sum_{j=1}^{m}\sum_{i=1}^{n}A_{i}\{1-Y_{i}(t)\}\left[\bone(T_{i}^{*(j)}\geq t)-S_{1}\{t\mid H_{i}(t);\hat{\theta}\}\right]+\frac{n^{1/2}}{n_{1}}\sum_{i=1}^{n}(1-A_{i})\phi_{10,i}(t)\\
 &  & +\frac{n^{1/2}}{n_{1}}\sum_{i=1}^{n}A_{i}\left[\phi_{11,i}(t)+Y_{i}(t)+\{1-Y_{i}(t)\}(1-I_{i})S_{1}\{t\mid H_{i}(t);\theta\}-S_{1}^{\delta\text{-cb}}(t)\right]+o_{p}(1),
\end{eqnarray*}
where 
\begin{eqnarray}
\phi_{11,i}(t) & = & \left[\left\{ \tilde{g}_{1,2}(t)-\tilde{g}_{1,1}(t)\right\} ^{\T}\Gamma_{1}^{-1}H_{1,i}-\int_{U_{i}}^{t}\frac{\tilde{g}_{1,0}(u)}{s_{0}(\beta_{1},u)}\de M_{1,i}(u)\right]\label{eq:tilde-phi1}\\
\phi_{10,i}(t) & = & \left[\left\{ \tilde{g}_{0,2}(t)-\tilde{g}_{0,1}(t)\right\} ^{\T}\Gamma_{0}^{-1}H_{0,i}-\int_{U_{i}}^{t}\frac{\tilde{g}_{0,0}(u)}{s_{0}(\beta_{0},u)}\de M_{0,i}(u)\right].\label{eq:tilde-phi2}
\end{eqnarray}

Because the imputation mechanism for the censored control subjects
is the same, the martingale  representation for $\hat{S}_{0,\mi}(t)$
remains the same as in (\ref{eq:S0}). Finally, we can decompose $\hat{\Delta}_{\tau,\mi}$
by the martingale representation 
\[
n^{1/2}(\hat{\Delta}_{\tau,\mi}-\Delta_{\tau}^{\delta\text{-bc}})=\sum_{k=1}^{(1+m)n}\xi_{n,k}+o_{p}(1),
\]
where the $\xi_{n,k}$ terms are given in (\ref{eq:xi1}) with $\phi_{11,i}(t)$,
$\phi_{10,i}(t)$, and $\phi_{0,i}(t)$ given in (\ref{eq:tilde-phi1}),
(\ref{eq:tilde-phi2}) and (\ref{eq:phi0}), respectively.

For both the $\delta$-adjusted and control-based Cox models, it follows
by the martingale CLT, $n^{1/2}\left(\hat{\Delta}_{\tau,\mi}-\Delta_{\tau}^{\sen}\right)$
converges to a Normal distribution with mean zero and a finite variance
\begin{equation}
V_{\tau,\mi}^{\sen}=\sum_{k=1}^{(1+m)n}\E(\xi_{n,k}^{2}\mid\F_{n,k-1})=\sum_{a=0}^{1}\left(\sigma_{a,1}^{2}+\sigma_{a,2}^{2}\right),\label{eq:sigma}
\end{equation}
where $\sen$ denotes either $\delta$-adj or $\delta$-cb, and 
\begin{eqnarray*}
\sigma_{0,1}^{2} & = & \frac{1}{p_{0}}\E\left(\left[\int_{0}^{\tau}\psi_{0}(t)\{(1-A_{i})[\phi_{10,i}(t)+\phi_{0,i}(t)+Y_{i}(t)\right.\right.\\
 &  & +\left.\left.\{1-Y_{i}(t)\}(1-I_{i})S_{a}\{t\mid H_{i}(t);\theta\}]-S_{a}^{\sen}(t)\}\de t\right]^{2}\right)\\
\sigma_{1,1}^{2} & = & \frac{1}{p_{1}}\E\left\{ \left(\int_{0}^{\tau}\psi_{1}(t)A_{i}[\phi_{11,i}(t)+Y_{i}(t)\right.\right.\\
 &  & +\left.\left.\vphantom{\int_{a}^{a}}\{1-Y_{i}(t)\}(1-I_{i})S_{a}\{t\mid H_{i}(t);\theta\}-S_{a}^{\sen}(t)]\de t\right)\right\} ,\\
\sigma_{a,2}^{2} & = & \frac{1}{p_{a}m}\V\left[\int_{0}^{\tau}\psi_{a}(t)\bone(A_{i}=a)\{1-Y_{i}(t)\}\{\bone(T_{i}^{*(j)}\geq t)-S_{a}\{t\mid H_{i}(t);\theta\}\}\de t\right],
\end{eqnarray*}
for $a=0,1$.

\section{Proof of Theorem \ref{th:WBconsistency}\label{sec:Proof-of-TheoremWB}}

We provide the proof of Theorem \ref{th:WBconsistency}, which draws
on the martingale central limit theory \citep{hall1980martingale}
and the asymptotic property of weighted sampling of martingale difference
arrays \citep{pauly2011weighted}. 

First, by the law of large numbers, we have {\small{}
\begin{eqnarray*}
 &  & \sum_{k=1}^{n_{1}}\xi_{n,k}^{2}\\
 & = & \frac{n}{n_{1}^{2}}\sum_{i=1}^{n_{1}}\left(\int_{0}^{\tau}\psi_{1}(t)A_{i}\left[\phi_{11,i}(t)+Y_{i}(t)+\{1-Y_{i}(t)\}(1-I_{i})S_{1}\{t\mid H_{i}(t);\theta\}-S_{1}^{\sen}(t)\right]\de t\right)^{2}\\
 & \overset{p}{\to} & \frac{1}{p_{1}}\E\left\{ \left(\int_{0}^{\tau}\psi_{1}(t)A_{i}\left[\phi_{11,i}(t)+Y_{i}(t)+\{1-Y_{i}(t)\}(1-I_{i})S_{1}\{t\mid H_{i}(t);\theta\}-S_{1}^{\sen}(t)\right]\de t\right)^{2}\right\} \\
 & = & \sigma_{1,1}^{2},
\end{eqnarray*}
}and {\small{}
\begin{eqnarray*}
 &  & \sum_{k=n_{1}+1}^{(1+m)n_{1}}\xi_{n,k}^{2}\\
 & = & \frac{n}{n_{1}^{2}}\sum_{i=1}^{n_{1}}\frac{1}{m^{2}}\sum_{j=1}^{m}\left[\int_{0}^{\tau}\psi_{1}(t)A_{i}\{1-Y_{i}(t)\}[\bone(T_{i}^{*(j)}\geq t)-S_{1}\{t\mid H_{i}(t);\hat{\theta}\}]\de t\right]^{2}\\
 & \overset{p}{\to} & \frac{1}{p_{1}m}\E\left(\var\left[\int_{0}^{\tau}\psi_{1}(t)A_{i}\{1-Y_{i}(t)\}[\bone(T_{i}^{*(j)}\geq t)-S_{1}\{t\mid H_{i}(t);\hat{\theta}\}]\de t\mid O_{1:n}\right]\right)\\
 & = & \sigma_{1,2}^{2},
\end{eqnarray*}
}as $n\rightarrow\infty.$ Similarly, by the law of large numbers,
we have $\sum_{k=(1+m)n_{1}+1}^{(1+m)n_{1}+n_{0}}\xi_{n,k}^{2}\overset{p}{\to}\sigma_{0,1}^{2}$,
and $\sum_{k=(1+m)n_{1}+n_{0}+1}^{(1+m)n}\xi_{n,k}^{2}\overset{p}{\to}\sigma_{0,2}^{2}$.
Therefore, we have 
\begin{equation}
\sum_{k=1}^{(1+m)n}\xi_{n,k}^{2}\overset{p}{\to}V_{\tau,\mi}^{\sen},\label{eq:cond1}
\end{equation}
as $n\rightarrow\infty.$

Second, we show 
\begin{equation}
\underset{1\leq k\leq(1+m)n}{\max}|\xi_{n,k}|\overset{p}{\to}0,\label{eq:cond2}
\end{equation}
as $n\rightarrow\infty$. Toward this end, for any $\epsilon>0$,
\begin{eqnarray*}
\bP\left(\underset{1\leq k\leq n_{1}}{\max}\vert\xi_{n,k}\vert>\epsilon\right) & \leq & n_{1}\bP\left(\vert\xi_{n,k}\vert>\epsilon\right)=n_{1}\bP\left(\xi_{n,k}^{4}>\epsilon^{4}\right)\\
 & \leq & \frac{n^{2}}{n_{1}^{3}\epsilon^{4}}\E\left(\int_{0}^{\tau}\psi_{1}(t)A_{i}\left[S_{1}\{t\mid H_{i}(t);\hat{\theta}\}-S_{1}^{\sen}(t)\right]\de t\right)^{4}\rightarrow0,
\end{eqnarray*}
where the second inequality follows from the Markov inequality, and
the convergence follows because the expectation term is bounded due
to the natural range of the survival functions. Similarly, we have
\[
\bP\left(\underset{n_{1}+1\leq k\leq(1+m)n_{1}}{\max}\vert\xi_{n,k}\vert>\epsilon\right)\leq\frac{n^{2}}{n_{1}^{3}m^{3}\epsilon^{4}}\E\left\{ \int_{0}^{\tau}\psi_{1}(t)A_{i}[\bone(T_{i}^{*(j)}\geq t)-S_{1}\{t\mid H_{i}(t);\hat{\theta}\}]\de t\right\} ^{4}\rightarrow0,
\]
as $n\rightarrow\infty$. Therefore, $\bP(\underset{1\leq k\leq(1+m)n_{1}}{\max}\vert\xi_{n,k}\vert>\epsilon)\rightarrow0$,
as $n\rightarrow\infty$. Similarly, $\bP(\underset{(1+m)n_{1}+1\leq k\leq(1+m)n}{\max}\vert\xi_{n,k}\vert>\epsilon)\rightarrow0$,
as $n\rightarrow\infty$. Then (\ref{eq:cond2}) holds.

Third, we show 
\begin{equation}
\underset{n}{\sup}\E\left(\underset{1\leq k\leq(1+m)n}{\max}\xi_{n,k}^{2}\right)<\infty.\label{eq:cond3}
\end{equation}
For any $n$, by Assumption \ref{asump:consistency}, 
\begin{eqnarray*}
\E\left(\underset{1\leq k\leq n_{1}}{\max}\xi_{n,k}^{2}\right) & \leq & \E\left(n_{1}\xi_{n,k}^{2}\right)\\
 & = & \frac{n}{n_{1}}\E\left(\int_{0}^{\tau}\left[\psi_{1}(t)A_{i}S_{1}\{t\mid H_{i}(t);\hat{\theta}\}-S_{1}^{\sen}(t)\right]\de t\right)^{2}<\infty,
\end{eqnarray*}
and 
\begin{eqnarray*}
\E\left(\underset{n_{1}+1\leq k\leq(1+m)n_{1}}{\max}\xi_{n,k}^{2}\right) & \leq & \E\left(nm\xi_{n,k}^{2}\right)\\
 & = & \frac{n}{mn_{1}}\E\left(\int_{0}^{\tau}\psi_{1}(t)A_{i}[\bone(T_{i}^{*(j)}\geq t)-S_{1}\{t\mid H_{i}(t);\hat{\theta}\}]\de t\right)^{2}<\infty.
\end{eqnarray*}
Therefore, $\E(\max_{1\leq k\leq(1+m)n_{1}}\xi_{n,k}^{2})\leq\E(\max_{1\leq k\leq n_{1}}\xi_{n,k}^{2})+\E(\max_{n_{1}+1\leq k\leq(1+m)n_{1}}\xi_{n,k}^{2})<\infty$.
Similarly, $\E(\max_{n_{1}(1+m)+1\leq k\leq n(1+m)}\xi_{n,k}^{2})<\infty$.
Then (\ref{eq:cond3}) follows.

Given the results in (\ref{eq:cond1}) and (\ref{eq:cond2}), the
martingale CLT implies that 
\[
\sum_{k=1}^{(1+m)n}\xi_{n,k}\overset{d}{\to}\mathcal{N}(0,V_{\tau,\mi}^{\sen}),
\]
as $n\rightarrow\infty.$ Given the results in (\ref{eq:cond1}),
(\ref{eq:cond2}), and (\ref{eq:cond3}), Theorem 2.1 in \citet{pauly2011weighted}
yields 
\begin{equation}
\underset{r}{\text{sup}}\left|\bP\left\{ \{(1+m)n\}^{1/2}\left.\sum_{k=1}^{(1+m)n}\frac{u_{k}}{\{n(1+m)\}^{1/2}}\xi_{n,k}\leq r\large\right\vert O_{1:n}\right\} -\Phi\left(\frac{r}{\sigma}\right)\right|\overset{p}{\to}0,\label{eq:ori_converge}
\end{equation}
as $n\rightarrow\infty$, where $\Phi(\cdot)$ denotes the cumulative
distribution function of the standard normal distribution.

Let $W_{L}=n^{-1/2}\sum_{k=1}^{(1+m)n}\xi_{n,k}u_{k}$. By Theorem
\ref{th:MI-point} and (\ref{eq:ori_converge}), we have 
\begin{equation}
\underset{r}{\text{sup}}\left|\bP\left(n^{1/2}W_{L}\leq r\mid O_{1:n}\right)-\bP\left\{ n^{1/2}\left(\hat{\Delta}_{\tau,\mi}-\Delta_{\tau}^{\sen}\right)\leq r\right\} \right|\overset{p}{\rightarrow}0,\label{eq:ori_mart}
\end{equation}
as $n\rightarrow\infty$.

Lastly, to prove Theorem \ref{th:WBconsistency}, it remains to show
that 
\begin{equation}
\bP\left\{ n^{1/2}(W_{L}-W_{L}^{*})\mid O_{1:n}\right\} \overset{p}{\to}0,\label{eq:residual}
\end{equation}
as $n\rightarrow\infty$. To unify the notation for both treatment
group, define $\Phi_{1,i}(t)=\phi_{11,i}(t)$, $\Phi_{0,i}(t)=\phi_{10,i}(t)+\phi_{0,i}(t)$,
$\hat{\Phi}_{1,i}(t)=\hat{\phi}_{11,i}(t)$, and $\hat{\Phi}_{0,i}(t)=\hat{\phi}_{10,i}(t)+\hat{\phi}_{0,i}(t)$.
The difference between $W_{L}$ and $W_{L}^{*}$ can be decomposed
to six parts, 
\[
n^{1/2}(W_{L}-W_{L}^{*})=\sum_{k=1}^{n(1+m)}n^{-1/2}u_{k}(n^{1/2}\hat{\xi}_{n,k}-n^{1/2}\xi_{n,k})=\sum_{a=0}^{1}\sum_{l=1}^{3}R_{al,n},
\]
where 
\begin{eqnarray*}
R_{a1,n} & = & \sum_{i=1}^{n}\frac{n^{1/2}}{n_{a}}u_{i}\bone(A_{i}=a)\int_{0}^{\tau}\psi_{a}(t)\left\{ \hat{S}_{a,\mi}(t)-S_{a}^{\sen}(t)\right\} \de t,\\
R_{a2,n} & = & \sum_{i=1}^{n}\frac{n^{1/2}}{n_{a}}u_{i}\bone(A_{i}=a)\int_{0}^{\tau}\psi_{a}(t)\left\{ \hat{\Phi}_{a,i}(t)-\Phi_{a,i}(t)\right\} \de t,\\
R_{a3,n} & = & \sum_{i=1}^{n}\frac{n^{1/2}}{n_{a}}u_{i}\bone(A_{i}=a)\\
 &  & \times\int_{0}^{\tau}\psi_{a}(t)\{1-Y_{i}(t)\}(1-I_{i})\left[S_{a}\{t\mid H_{i}(t);\hat{\theta}\}-S_{a}\{t\mid H_{i}(t);\theta\}\right]\de t,
\end{eqnarray*}
for $a=0,1$.

Given that the bootstrap weights satisfy $\E(u_{k}^{2}\mid O_{1:n})=1$,
we have 
\begin{eqnarray*}
\E\left(R_{a1,n}^{2}\vert O_{1:n}\right) & = & \frac{n}{n_{a}^{2}}n_{a}\E(u_{i}^{2})\left[\int_{0}^{\tau}\psi_{a}(t)\left\{ \hat{S}_{a,\mi}(t)-S_{a}^{\sen}(t)\right\} \de t\right]^{2}\\
 & = & \frac{n}{n_{a}}\left[\int_{0}^{\tau}\psi_{a}(t)\left\{ \hat{S}_{a,\mi}(t)-S_{a}^{\sen}(t)\right\} \de t\right]^{2}\overset{p}{\to}0,
\end{eqnarray*}
as $n\rightarrow\infty$, for $a=0,1$. Also, we have 
\[
\E\left(R_{a2,n}^{2}\vert O_{1:n}\right)=\frac{n}{n_{a}^{2}}\sum_{i=1}^{n}\bone(A_{i}=a)\left[\int_{0}^{\tau}\psi_{a}(t)\left\{ \hat{\Phi}_{a,i}(t)-\Phi_{a,i}(t)\right\} \de t\right]^{2}\overset{p}{\to}0,
\]
as $n\rightarrow\infty$, for $a=0,1$, where the convergence follows
by Assumption \ref{asump:consistency} and the results in Section
\ref{sec:Preliminary}. Similarly, we have 
\begin{multline*}
\E\left(R_{a3,n}^{2}|O_{1:n}\right)\\
=\frac{n}{n_{a}^{2}}\sum_{i=1}^{n_{a}}\bone(A_{i}=a)\left[\int_{0}^{\tau}\psi_{a}(t)\{1-Y_{i}(t)\}(1-I_{i})\right.\left.\left\{ S_{a}(t\mid O_{i};\hat{\theta}_{a})-S_{a}\{t\mid H_{i}(t);\theta\}\right\} \de t\right]^{2}\overset{p}{\to}0,
\end{multline*}
as $n\rightarrow\infty$, for $a=0,1$. Therefore, for any $\epsilon>0$,
\[
\bP\{|R_{a1,n}|>\epsilon\mid O_{1:n}\}\overset{p}{\to}0,\quad\bP\{|R_{a2,n}|>\epsilon\mid O_{1:n}\}\overset{p}{\to}0,\quad\bP\{|R_{a3,n}|>\epsilon\mid O_{1:n}\}\overset{p}{\to}0,
\]
as $n\rightarrow\infty$, for $a=0,1$. Then we obtain (\ref{eq:residual}).
The conclusion of Theorem \ref{th:WBconsistency} follows.

% Preview body

\section{ Simulation study\label{sec:sim}}

We conduct simulation studies to evaluate the finite sample performance
of the proposed SMIM framework. For illustration, we focus on the
$\delta$-adjusted and control-based models for sensitivity analysis
and the RMST as the treatment effect estimand. We start with a simple setup with one covariate in Section \ref{sec:sim1}
and then consider a setting motivated by the ACTG175 trial data in Section \ref{sec:sim2}.

\subsection{ Simulation one: a simple setup\label{sec:sim1}}
For both the treatment
and control groups, each with sample size $n\in\{500,1000\}$, the
confounder is generated by $X_{i}\sim\mathcal{N}(0,1)$. In the treatment
group, $T$ follows the Cox model with the hazard rate $\lambda_{1}(t\mid X_{i})=\lambda_{1}(t)\exp(\beta_{1}X_{i})$,
where $\lambda_{1}(t)=0.35$ and $\beta_{1}=0.75$. We consider censoring
due to the end of the study and premature dropout. We generate the
censoring time to dropout, $C_{i}$, according to a Cox model with
the hazard rate $\lambda_{C}(t\mid X_{i})=\lambda_{C}(t)\exp(\beta_{C}X_{i})$,
where $\lambda_{C}(t)=0.15$ and $\beta_{C}=0.75$. The maximum follow
up time is $L=3.25$. The observed time is $U_{i}=T_{i}\land C_{i}\land L$.
If $U_{i}=T_{i}$, the event indicator is $I_{i}=1$; if $U_{i}=L$,
then $I_{i}=0$ and the censoring type is $R_{i}=1$; if $U_{i}=C_{i}$,
then $I_{i}=0$ and the censoring type is $R_{i}=2$. Under the data
generating mechanism, the average percentages of $I_{i}=1$, $R_{i}=1$,
and $R_{i}=2$ are $53\%$, $25\%$, and $22\%$, respectively. In
the control group, $T_{i}$ follows the hazard rate $\lambda_{0}(t\mid X)=\lambda_{0}(t)\exp(\beta_{0}X_{i})$,
where $\lambda_{0}(t)=0.40$ and $\beta_{0}=0.75$. The censoring
time $C_{i}$ follows the same model as in the treatment group. For
the dropout subjects with $R_{i}=2$ in treatment group, the hazard
rate for events after censoring are $\delta\lambda_{1}(t)\exp(\beta_{1}X_{i})$
for delta-adjusted model and $\lambda_{0}(t)\exp(\beta_{0}X_{i})$
for control-based models. For the dropout subjects with $R_{i}=2$
in control group, the hazard rate for event after censoring remains
the same, which correspondsto the case when the control treatment
is a placebo or the standard of care. The true RMST estimand under
the $\delta$-adjusted model is $\Delta_{\tau}^{\delta\text{-adj}}=\mu_{1,\tau}^{\delta\text{-adj}}-\mu_{0,\tau}$
with $\tau=3$. We assess the proposed method to implement the sensitivity
analysis for the treatment group when the true parameter $\delta$
is $1.5$, while the analysis parameter $\delta$ varies in a pre-specified
set $\{0.5,1,1.5,2,2.5\}$. The true RMST estimand under the control-based
model are $\Delta_{\tau}^{\text{control-adj}}$ with $\tau=3$. 

We use MI for imputing the censored event times following Steps MI-1-1,
MI-1-2 and MI-1-3 in Section \ref{sec:D-CBimpM} with imputation size
$m\in\{10,\,20,\,50\}$. We compare the standard MI inference and
the proposed wild bootstrap inference. For the standard MI inference,
the $100(1-\alpha)\%$ confidence intervals are calculated as $(\hat{\Delta}_{\tau,\mi}-z_{1-\alpha/2}\hat{V}_{\mi}^{1/2},\hat{\Delta}_{\tau,\mi}+z_{1-\alpha/2}\hat{V}_{\mi}^{1/2})$,
where $z_{1-\alpha/2}$ is the $(1-\alpha/2)$th quantile of the standard
normal distribution. For the proposed wild bootstrap procedure, we
sample the weights $\mu_{k}$ from the standard normal distribution,
and calculate the variance estimate $\hat{V}_{\text{WB}}$ based on
$100$ replications. The corresponding $100(1-\alpha)\%$ confidence
intervals are calculated as $(\hat{\Delta}_{\tau,\mi}-z_{1-\alpha/2}\hat{V}_{\text{WB}}^{1/2},\hat{\Delta}_{\tau,\mi}+z_{1-\alpha/2}\hat{V}_{\text{WB}}^{1/2})$.
We assess the performance in terms of the relative bias of the variance
estimator and the coverage rate of confidence intervals. The relative
bias of the variance estimators are calculated as $\{\E(\hat{V}_{\mi}^{1/2})-\V(\hat{\Delta}_{\tau,\mi}^{1/2})\}/\V(\hat{\Delta}_{\tau,\mi}^{1/2})\times100\%$
and $\{\E(\hat{V}_{\text{WB}}^{1/2})-\V(\hat{\Delta}_{\tau,\mi}^{1/2})\}/\V(\hat{\Delta}_{\tau,\mi}^{1/2})\times100\%$.
The coverage rate of the $100(1-\alpha)\%$ confidence intervals is
estimated by the percentage of the Monte Carlo samples for which the
confidence intervals contain the true value.

Table \ref{t:CNAR} presents the simulation results for the sensitivity
analysis of $\delta$-adjusted estimand $\Delta_{\tau}^{\delta\text{-adj}}$
based on 1000 Monte Carlo samples. When the imputation model is correctly
specified with $\delta=1.5$, the MI point estimator $\hat{\Delta}_{\tau,\mi}$
is unbiased of the true estimand $\Delta_{\tau}^{\delta\text{-adj}}$.
When the analysis sensitivity parameter is lower (higher) than the
true parameter $\delta=1.5$, the MI point estimator produces higher
(lower) RMST for the treatment group, and therefore $\hat{\Delta}_{\tau,\mi}$
is biased upward (downward). When the true sensitivity parameter is
correctly specified, Rubin's combining rule overestimates the true
standard deviation with the relative bias ranging from $7.0\%$ to
$12.2\%$; consequently, the coverage rates are larger than the nominal
level $95\%$. In contrast, our proposed wild bootstrap procedure
is unbiased; as a result, the coverage rates of the confidence intervals
are close to the nominal level. Moreover, the proposed method is not
sensitive to the number of imputations $m$. We observed similar behavior
for the sensitivity analysis of control-based models for sensitivity
analysis and summarized in Table \ref{t:CBT}.

\begin{table}[ht]
\centering \caption{Simulation results for the true estimand $\Delta_{\tau}^{\delta\text{-adj}}=0.054$
with the true sensitivity parameter $\delta=1.5$: point estimate,
true standard deviation, relative bias of the standard error estimator,
coverage of interval estimate using Rubin's method and the proposed
wild bootstrap method}
\label{t:CNAR} \resizebox{\textwidth}{!}{%
\begin{tabular}{cclcccccccc}
\hline 
 &  &  &  &  & \multicolumn{2}{c}{Standard error} & \multicolumn{2}{l}{Relative Bias} & \multicolumn{2}{l}{Coverage (\%)}\tabularnewline
 &  &  & Point est & True sd & \multicolumn{2}{c}{($\times10^{2}$)} & \multicolumn{2}{c}{(\%)} & \multicolumn{2}{c}{for 95\% CI}\tabularnewline
n & m & Model & ($\times10^{2}$) & ($\times10^{2}$) & Rubin$^{a}$ & WB & Rubin$^{a}$ & WB & Rubin$^{a}$ & WB\tabularnewline
500 & 10 & $\delta=$0.50 & 15.8 & 6.93 & 7.43 & 6.78 & 7.24 & -2.18 & 71.0 & 66.2\tabularnewline
 &  & $\delta=$1.00 & 9.3 & 6.91 & 7.41 & 6.74 & 7.31 & -2.43 & 94.3 & 90.8\tabularnewline
 &  & $\delta=$1.50 & 5.0 & 6.89 & 7.38 & 6.74 & 7.11 & -2.15 & \textbf{97.0} & \textbf{95.1}\tabularnewline
 &  & $\delta=$2.00 & 2.0 & 6.87 & 7.35 & 6.75 & 6.94 & -1.74 & 94.5 & 92.0\tabularnewline
 &  & $\delta=$2.50 & -0.3 & 6.85 & 7.32 & 6.77 & 6.84 & -1.30 & 89.3 & 85.7\tabularnewline
 & 20 & $\delta=$0.50 & 15.8 & 6.92 & 7.41 & 6.76 & 7.12 & -2.28 & 71.3 & 65.5\tabularnewline
 &  & $\delta=$1.00 & 9.3 & 6.90 & 7.39 & 6.73 & 7.14 & -2.53 & 93.9 & 90.3\tabularnewline
 &  & $\delta=$1.50 & 5.1 & 6.88 & 7.36 & 6.73 & 6.99 & -2.22 & \textbf{96.6} & \textbf{94.9}\tabularnewline
 &  & $\delta=$2.00 & 2.0 & 6.86 & 7.33 & 6.74 & 6.89 & -1.76 & 94.4 & 91.9\tabularnewline
 &  & $\delta=$2.50 & -0.3 & 6.84 & 7.31 & 6.75 & 6.84 & -1.28 & 89.4 & 86.0\tabularnewline
 & 50 & $\delta=$0.50 & 15.8 & 6.90 & 7.41 & 6.75 & 7.37 & -2.07 & 71.3 & 65.6\tabularnewline
 &  & $\delta=$1.00 & 9.3 & 6.88 & 7.38 & 6.72 & 7.38 & -2.32 & 94.1 & 91.0\tabularnewline
 &  & $\delta=$1.50 & 5.0 & 6.86 & 7.35 & 6.72 & 7.22 & -2.01 & \textbf{96.6} & \textbf{95.0}\tabularnewline
 &  & $\delta=$2.00 & 2.0 & 6.84 & 7.32 & 6.73 & 7.09 & -1.56 & 94.7 & 91.7\tabularnewline
 &  & $\delta=$2.50 & -0.3 & 6.82 & 7.30 & 6.75 & 7.01 & -1.10 & 89.3 & 85.9\tabularnewline
 & N/A & Tian et.al. 2014 & 9.4 & 7.10 & - & 7.56 & - & 6.40 & - & 92.9\tabularnewline
\hline 
1000 & 10 & $\delta=$0.50 & 16.3 & 4.72 & 5.25 & 4.80 & 11.19 & 1.58 & 45.4 & 37.5\tabularnewline
 &  & $\delta=$1.00 & 9.8 & 4.68 & 5.24 & 4.77 & 11.87 & 1.98 & 87.9 & 84.2\tabularnewline
 &  & $\delta=$1.50 & 5.6 & 4.66 & 5.21 & 4.78 & 11.98 & 2.57 & \textbf{97.7} & \textbf{95.2}\tabularnewline
 &  & $\delta=$2.00 & 2.5 & 4.64 & 5.19 & 4.79 & 12.04 & 3.22 & 94.4 & 91.3\tabularnewline
 &  & $\delta=$2.50 & 0.2 & 4.62 & 5.18 & 4.80 & 12.14 & 3.85 & 85.2 & 80.9\tabularnewline
 & 20 & $\delta=$0.50 & 16.3 & 4.71 & 5.25 & 4.79 & 11.39 & 1.76 & 45.0 & 37.8\tabularnewline
 &  & $\delta=$1.00 & 9.8 & 4.67 & 5.23 & 4.77 & 12.02 & 2.08 & 87.9 & 84.6\tabularnewline
 &  & $\delta=$1.50 & 5.6 & 4.64 & 5.21 & 4.77 & 12.14 & 2.68 & \textbf{97.7} & \textbf{95.0}\tabularnewline
 &  & $\delta=$2.00 & 2.5 & 4.62 & 5.19 & 4.78 & 12.20 & 3.35 & 94.1 & 91.5\tabularnewline
 &  & $\delta=$2.50 & 0.2 & 4.61 & 5.17 & 4.79 & 12.28 & 3.97 & 85.7 & 81.8\tabularnewline
 & 50 & $\delta=$0.50 & 16.3 & 4.70 & 5.24 & 4.79 & 11.39 & 1.78 & 45.3 & 37.5\tabularnewline
 &  & $\delta=$1.00 & 9.8 & 4.66 & 5.22 & 4.76 & 12.06 & 2.13 & 88.0 & 84.6\tabularnewline
 &  & $\delta=$1.50 & 5.5 & 4.64 & 5.20 & 4.76 & 12.19 & 2.74 & \textbf{97.5} & \textbf{95.2}\tabularnewline
 &  & $\delta=$2.00 & 2.5 & 4.61 & 5.18 & 4.77 & 12.27 & 3.41 & 94.1 & 91.4\tabularnewline
 &  & $\delta=$2.50 & 0.2 & 4.60 & 5.17 & 4.78 & 12.34 & 4.03 & 85.4 & 81.7\tabularnewline
 & N/A & Tian et.al. 2014 & 9.9 & 4.90 & - & 5.35 & - & 9.28 & - & 88.2\tabularnewline
\hline 
\end{tabular}}
\end{table}

\begin{table}[ht]
\centering \caption{Simulation results for the true estimand $\Delta_{\tau}^{\text{control-adj}}=1.783$
based on control-based method: point estimate, true standard deviation,
relative bias of the standard error estimator, coverage of interval
estimate using Rubin's method and the proposed wild bootstrap method}
\label{t:CBT} \resizebox{\textwidth}{!}{ %
\begin{tabular}{clccccccccc}
\hline 
 &  &  &  &  & \multicolumn{2}{c}{Standard error} & \multicolumn{2}{c}{Relative Bias} & \multicolumn{2}{c}{Coverage (\%)}\tabularnewline
 &  &  & Point est & True sd & \multicolumn{2}{c}{($\times10^{2}$)} & \multicolumn{2}{c}{(\%)} & \multicolumn{2}{c}{for 95\% CI}\tabularnewline
n & Model & m & ($\times10^{2}$) & ($\times10^{2}$) & Rubin$^{a}$ & WB & Rubin$^{a}$ & WB & Rubin$^{a}$ & WB\tabularnewline
\hline 
500 & Control-based & 10 & 179.0 & 4.58 & 5.24 & 4.76 & 14.34 & 3.87 & \textbf{97.2} & \textbf{95.1}\tabularnewline
 &  & 20 & 179.0 & 4.58 & 5.22 & 4.75 & 13.95 & 3.62 & \textbf{97.4} & \textbf{95.2}\tabularnewline
 &  & 50 & 179.0 & 4.57 & 5.22 & 4.74 & 14.16 & 3.76 & \textbf{97.3} & \textbf{95.3}\tabularnewline
 & Tian et.al. 2014 & N/A & 184.6 & 4.81 & 5.34 & - & 10.93 & - & 80.2 & -\tabularnewline
\hline 
1000 & Control-based & 10 & 179.1 & 3.30 & 3.70 & 3.37 & 11.94 & 1.97 & \textbf{96.6} & \textbf{94.4}\tabularnewline
 &  & 20 & 179.1 & 3.30 & 3.69 & 3.36 & 12.10 & 2.03 & \textbf{96.5} & \textbf{94.2}\tabularnewline
 &  & 50 & 179.1 & 3.29 & 3.69 & 3.36 & 12.21 & 2.14 & \textbf{96.4} & \textbf{94.5}\tabularnewline
 & Tian et.al. 2014 & N/A & 184.8 & 3.53 & 3.78 & - & 7.08 & - & 61.1 & -\tabularnewline
\hline 
\end{tabular}}
\end{table}

\subsection{ Simulation two: ACTG175\label{sec:sim2}}
We consider a simulation setup that is similar to ACTG175
data. The confounder is generated by $X_{1i}\sim\mathcal{N}(0,1)$
and $X_{2i}\sim\text{Bernoulli}(0.15)$. In the treatment group, $T$
follows the Cox model with the hazard rate $\lambda_{1}(t\mid X_{1i}X_{2i})=\lambda_{1}(t)\exp(\beta_{1}X_{1i}+\beta_{2}X_{2i}),$
where $\lambda_{1}=0.03$, $\beta_{1}=0.24$ and $\beta_{2}=0.04$.
We consider censoring due to the end of the study and premature dropout.
We generate the censoring time to dropout, $C_{i}$, according to
a Cox model with the hazard rate $\lambda_{C}(t\mid X_{1i}X_{2i})=\lambda_{C}(t)\exp(\beta_{C1}X_{1i}+\beta_{C2}X_{2i})$,
where $\lambda_{C}(t)=0.01$, $\beta_{C1}=0.24$, $\beta_{C2}=0.20$.
The maximum follow up time is $L=40$. The observed time is $U_{i}=T_{i}\land C_{i}\land L$.
If $U_{i}=T_{i}$, the event indicator is $I_{i}=1$; if $U_{i}=L$,
then $I_{i}=0$ and the censoring type is $R_{i}=1$; if $U_{i}=C_{i},$then
$I_{i}=0$ and the censoring type is $R_{i}=2$. Under the data generating
mechanism, the average percentages of $I_{i}=1$, $R_{i}=1$, and
$R_{i}=2$ are $60\%$, $20\%$ and $20\%$, respectively.In the control
group, $T_{i}$ follows the hazard rate $\lambda_{0}(t\mid X_{1i},X_{2i})=\lambda_{0}(t)\exp(\beta_{01}X_{1i}+\beta_{02}X_{2i}),$
where $\lambda_{0}(t)=0.03$, $\beta_{01}=-0.55$ and $\beta_{02}=0.65$.
The censoring time $C_{i}$ follows the same model as in the treatment
group. For the dropout subjects with $R_{i}=2$ in treatment group,
the hazard rate for events after censoring are $\delta\lambda_{1}(t)\exp(\beta_{1}X_{1i}+\beta_{2}X_{2i})$
for delta-adjusted model and $\lambda_{0}(t)\exp(\beta_{01}X_{1i}+\beta_{02}X_{2i})$
for control-based models. For the dropout subjects with $R_{i}=2$
in control group, the hazard rate for event after censoring remains
the same, which correspondsto the case when the control treatment
is a placebo or the standard of care. The true RMST estimand under
the $\delta$-adjusted model is $\Delta_{\tau}^{\delta\text{-adj}}=\mu_{1,\tau}^{\delta\text{-adj}}-\mu_{0,\tau}$
with $\tau=24$. We assess the proposed method to implement the sensitivity
analysis for the treatment group when the true parameter $\delta$
is $2$, while the analysis parameter $\delta$ varies in a pre-specified
set $\{1,2,3,4,5\}$. The true RMST estimand under the control-based
model are $\Delta_{\tau}^{\text{control-adj}}$ with $\tau=24$. The
estimation procedure are the same as the first simulation study. The
simulation results is summarized in Table \ref{t:delta2} and Table
\ref{t:CBT2} with similar observation in the first simulation study.

\begin{table}[ht]
\centering \caption{Simulation results for the true estimand $\Delta_{\tau}^{\text{control-adj}}=0.513$
based on control-based method: point estimate, true standard deviation,
relative bias of the standard error estimator, coverage of interval
estimate using Rubin's method and the proposed wild bootstrap method}
\label{t:delta2} \resizebox{\textwidth}{!}{ %
\begin{tabular}{cclcccccccc}
\hline 
 &  &  &  &  & \multicolumn{2}{c}{Standard error} & \multicolumn{2}{l}{Relative Bias} & \multicolumn{2}{l}{Coverage (\%)}\tabularnewline
 &  &  & Point est & True sd & \multicolumn{2}{c}{($\times10^{2}$)} & \multicolumn{2}{c}{(\%)} & \multicolumn{2}{c}{for 95\% CI}\tabularnewline
n & m & Model & ($\times10^{2}$) & ($\times10^{2}$) & Rubin$^{a}$ & WB & Rubin$^{a}$ & WB & Rubin$^{a}$ & WB\tabularnewline
500 & 10 & $\delta=$1 & 84.9 & 55.9 & 60.8 & 58.3 & 8.74 & 4.25 & 93.8 & 92.1\tabularnewline
 &  & $\delta=$2 & 50.3 & 56.5 & 60.4 & 58.0 & 6.91 & 2.69 & 96.2 & \textbf{95.1}\tabularnewline
 &  & $\delta=$3 & 26.1 & 56.8 & 59.6 & 57.9 & 4.93 & 2.03 & 94.2 & 93.2\tabularnewline
 &  & $\delta=$4 & 8.2 & 56.8 & 59.0 & 58.3 & 3.79 & 2.53 & 90.0 & 88.8\tabularnewline
 &  & $\delta=$5 & -5.0 & 57.0 & 58.6 & 58.5 & 2.82 & 2.65 & 84.6 & 84.4\tabularnewline
 & 20 & $\delta=$1 & 86.7 & 54.2 & 60.6 & 55.3 & 11.83 & 2.09 & 93.4 & 90.7\tabularnewline
 &  & $\delta=$2 & 52.2 & 54.5 & 60.0 & 55.2 & 10.15 & 1.37 & 97.5 & \textbf{95.4}\tabularnewline
 &  & $\delta=$3 & 27.9 & 54.8 & 59.3 & 55.4 & 8.17 & 1.07 & 95.1 & 93.2\tabularnewline
 &  & $\delta=$4 & 10.1 & 54.9 & 58.9 & 55.5 & 7.15 & 1.03 & 89.6 & 86.5\tabularnewline
 &  & $\delta=$5 & -2.9 & 55.0 & 58.5 & 55.6 & 6.28 & 1.16 & 85.0 & 82.3\tabularnewline
 & 50 & $\delta=$1 & 85.4 & 54.4 & 60.4 & 53.3 & 11.02 & -1.96 & 94.3 & 89.4\tabularnewline
 &  & $\delta=$2 & 51.0 & 55.0 & 60.0 & 53.4 & 9.08 & -2.79 & 97.2 & \textbf{94.9}\tabularnewline
 &  & $\delta=$3 & 26.8 & 55.2 & 59.3 & 53.5 & 7.40 & -3.11 & 94.3 & 91.7\tabularnewline
 &  & $\delta=$4 & 9.2 & 55.3 & 58.7 & 53.6 & 6.16 & -3.04 & 90.9 & 86.6\tabularnewline
 &  & $\delta=$5 & -4.2 & 55.3 & 58.3 & 53.8 & 5.55 & -2.59 & 86.2 & 80.5\tabularnewline
 & N/A & Tian et.al. 2014 & 92.8 & 55.4 & - & 55.5 & - & 0.29 & - & 88.5\tabularnewline
\hline 
1000 & 10 & $\delta=$1 & 87.0 & 38.6 & 43.0 & 41.1 & 11.25 & 6.46 & 90.7 & 88.1\tabularnewline
 &  & $\delta=$2 & 52.4 & 38.8 & 42.7 & 41.0 & 10.10 & 5.73 & 97.2 & \textbf{96.6}\tabularnewline
 &  & $\delta=$3 & 28.1 & 38.7 & 42.0 & 41.2 & 8.47 & 6.48 & 93.3 & 92.5\tabularnewline
 &  & $\delta=$4 & 10.6 & 38.7 & 41.7 & 41.3 & 7.59 & 6.56 & 84.9 & 84.2\tabularnewline
 &  & $\delta=$5 & -3.0 & 38.8 & 41.4 & 41.3 & 6.80 & 6.58 & 76.4 & 76.0\tabularnewline
 & 20 & $\delta=$1 & 85.4 & 39.5 & 42.8 & 39.1 & 8.46 & -0.93 & 90.2 & 86.5\tabularnewline
 &  & $\delta=$2 & 50.8 & 39.8 & 42.6 & 38.9 & 7.09 & -2.20 & 96.5 & \textbf{95.2}\tabularnewline
 &  & $\delta=$3 & 26.2 & 40.1 & 42.0 & 39.1 & 4.82 & -2.34 & 91.4 & 88.7\tabularnewline
 &  & $\delta=$4 & 8.5 & 40.1 & 41.6 & 39.3 & 3.71 & -2.03 & 82.4 & 78.5\tabularnewline
 &  & $\delta=$5 & -4.9 & 40.1 & 41.3 & 39.3 & 2.96 & -2.05 & 72.4 & 69.4\tabularnewline
 & 50 & $\delta=$1 & 86.8 & 39.1 & 42.7 & 37.8 & 9.19 & -3.22 & 88.1 & 83.2\tabularnewline
 &  & $\delta=$2 & 52.3 & 39.5 & 42.4 & 37.6 & 7.30 & -4.67 & 96.3 & \textbf{93.9}\tabularnewline
 &  & $\delta=$3 & 28.0 & 39.7 & 41.8 & 37.9 & 5.38 & -4.60 & 92.7 & 89.3\tabularnewline
 &  & $\delta=$4 & 10.4 & 39.8 & 41.4 & 37.9 & 4.22 & -4.71 & 83.6 & 78.3\tabularnewline
 &  & $\delta=$5 & -2.9 & 39.8 & 41.2 & 38.1 & 3.37 & -4.44 & 74.8 & 69.8\tabularnewline
 & N/A & Tian et.al. 2014 & 93.0 & 39.6 & - & 39.3 & - & -0.87 & - & 81.3\tabularnewline
\hline 
\end{tabular}}
\end{table}

\begin{table}[ht]
\centering \caption{Simulation results for the true estimand $\Delta_{\tau}^{\text{control-adj}}=0.843$
based on control-based method: point estimate, true standard deviation,
relative bias of the standard error estimator, coverage of interval
estimate using Rubin's method and the proposed wild bootstrap method}
\label{t:CBT2} \resizebox{\textwidth}{!}{ %
\begin{tabular}{clccccccccc}
\hline 
 &  &  &  &  & \multicolumn{2}{c}{Standard error} & \multicolumn{2}{c}{Relative Bias} & \multicolumn{2}{c}{Coverage (\%)}\tabularnewline
 &  &  & Point est  & True sd  & \multicolumn{2}{c}{($\times10^{2}$)} & \multicolumn{2}{c}{(\%)} & \multicolumn{2}{c}{for 95\% CI}\tabularnewline
n  & Model  & m  & ($\times10^{2}$)  & ($\times10^{2}$)  & Rubin$^{a}$  & WB  & Rubin$^{a}$  & WB  & Rubin$^{a}$  & WB\tabularnewline
\hline 
500 & Control-based & 10 & 85.2 & 55.1 & 60.6 & 58.1 & 9.90 & 5.36 & 96.8 & \textbf{95.7}\tabularnewline
 &  & 20 & 84.4 & 53.8 & 60.3 & 55.2 & 12.03 & 2.53 & 97.3 & \textbf{95.3}\tabularnewline
 &  & 50 & 87.1 & 53.2 & 60.2 & 53.5 & 13.08 & 0.55 & 97.0 & \textbf{94.7}\tabularnewline
 & Tian et.al. 2014 & N/A & 93.9 & 54.4 & - & 55.5 & - & 1.95 & - & 95.2\tabularnewline
\hline 
1000 & Control-based & 10 & 86.1 & 38.9 & 42.8 & 41.1 & 10.16 & 5.81 & 96.8 & \textbf{96.3}\tabularnewline
 &  & 20 & 83.5 & 38.7 & 42.7 & 39.1 & 10.53 & 1.13 & 96.3 & \textbf{95.3}\tabularnewline
 &  & 50 & 86.6 & 38.1 & 42.5 & 37.9 & 11.56 & -0.70 & 96.2 & \textbf{94.6}\tabularnewline
 & Tian et.al. 2014 & N/A & 93.1 & 39.3 & - & 39.3 & - & -0.10 & - & 94.2\tabularnewline
\hline 
\end{tabular}} 
\end{table}

\end{document}